\documentclass[lettersize,journal]{IEEEtran}

\usepackage{xcolor}
\usepackage{acronym}
\usepackage{graphicx}
\usepackage{amsmath}
\usepackage{tabularx,booktabs}
\usepackage{multicol}
\usepackage{makecell}
\usepackage{amsfonts}
\usepackage{bm}
\usepackage{tablefootnote}
\usepackage{subcaption}
\usepackage[hyphens]{url}
\usepackage{caption}
\usepackage{amssymb}
\usepackage{amsthm}
\usepackage{chngcntr}
\usepackage[table]{xcolor}
\usepackage{array}        
\usepackage{array, multirow}
\usepackage{caption}      
\usepackage{graphicx}
\usepackage{diagbox}
\usepackage[top=1.2cm,bottom=1.2cm,left=1.9cm,right=1.9cm]{geometry}
\usepackage{float} 
\usepackage{placeins}
\usepackage{tikz}
\usepackage{graphicx} 
\usepackage[caption=false,font=normalsize,labelfont=sf,textfont=sf]{subfig}
\usetikzlibrary{arrows.meta, positioning, shapes.geometric}
\usepackage{array}
\newcolumntype{C}{>{\centering\arraybackslash}X}
\usepackage[table]{xcolor}

\definecolor{better}{rgb}{0.7, 0.85, 1}
\definecolor{worse}{rgb}{1, 0.9, 0.6}    
\definecolor{neutral}{rgb}{1, 1, 1}
\definecolor{diaggray}{gray}{0.85}
\newcolumntype{Y}{>{\centering\arraybackslash}X}
\newcolumntype{?}{!{\vrule width 1.5pt}}
\usepackage[numbers,sort&compress]{natbib}
\acrodef{PMF}[PMF]{probability mass function}
\acrodef{MAC}[MAC]{medium access control}
\acrodef{IRSA}[IRSA]{irregular repetition slotted ALOHA}
\acrodef{i.i.d.}[i.i.d.]{independent and identically distributed}
\acrodef{DL}[DL]{deep learning}
\acrodef{DNN}[DNN]{deep neural network}
\acrodef{BN}[BN]{Bayesian network}
\acrodef{MC}[MC]{Markov chain}
\acrodef{MAB}[MAB]{multi-armed bandit}
\acrodef{IoT}[IoT]{internet of things}
\acrodef{URLLC}[URLLC]{ultra-reliable low-latency communication}
\acrodef{MARL}[MARL]{multi-agent reinforcement learning}
\acrodef{RL}[RL]{reinforcement learning}
\acrodef{mMTC}[mMTC]{massive machine-type communication}
\acrodef{CI}[CI]{confidence interval}
\acrodef{MDT}[MDT]{minimization of drive test}
\acrodef{KDE}[KDE]{kernel density estimation}
\acrodef{G-KDE}[G-KDE]{Gaussian kernel density estimation}
\acrodef{GAN}[GAN]{generative adversarial network}
\acrodef{VAE}[VAE]{variational autoencoder}
\acrodef{NF}[NF]{normalizing flow}
\acrodef{GMM}[GMM]{Gaussian mixture model}
\acrodef{KS}[KS]{Kolmogorov-Smirnov}
\acrodef{PDF}[PDF]{probability density function}
\acrodef{ERM}[ERM]{empirical risk minimization}
\acrodef{MNO}[MNO]{mobile network operator}
\acrodef{GPR}[GPR]{Gaussian process regression}
\acrodef{GPS}[GPS]{global positioning system}
\acrodef{RSRP}[RSRP]{reference signal received power}
\acrodef{SE}[SE]{squared exponential}
\acrodef{TTI}[TTI]{time transmission interval}
\acrodef{wKNN}[wKNN]{weighted k‐nearest‐neighbor}
\acrodef{OM}[O\&M]{operation \& maintenance}
\acrodef{3GPP}[3GPP]{3rd generation partnership project}
\acrodef{PCI}[PCI]{physical cell identity}
\acrodef{NLOS}[NLOS]{non-line-of-sight}
\acrodef{RQ}[RQ]{rational quadratic}
\acrodef{LTE}[LTE]{long term evolution}
\acrodef{RF}[RF]{random forest}
\acrodef{KNN}[KNN]{k‐nearest‐neighbor}
\acrodef{MLP}[MLP]{multilayer perceptron}
\acrodef{ReFlow}[ReFlow]{rectified flow}
\acrodef{CGAN}[CGAN]{conditional generative adversarial network}
\acrodef{NF}[NF]{normalizing flow}
\acrodef{UE}[UE]{user equipment}
\acrodef{RRM}[RRM]{radio resource management}
\acrodef{CRS}[CRS]{cell-specific reference signal}
\acrodef{RE}[RE]{resource element}
\acrodef{MAE}[MAE]{mean absolute error}
\acrodef{MedAE}[MedAE]{median absolute error}
\acrodef{KDE-KNN}[KDE-KNN]{kernel density estimation with k-nearest-neighbor}
\acrodef{KDE-GPR}[KDE-GPR]{kernel density estimation with Gaussian process regression}
\acrodef{KDE-RF}[KDE-RF]{kernel density estimation with random forest}
\acrodef{KDE-GPR(RQ)}[KDE-GPR(RQ)]{kernel density estimation with rational quadratic Gaussian process regression}
\acrodef{OFDM}[OFDM]{orthogonal frequency division multiplexing}
\acrodef{CSI}[CSI]{channel state information}
\acrodef{RBF}[RBF]{radial basis function} 
\acrodef{RT}[RT]{ray tracing}
\acrodef{1NN}[1NN]{1‐nearest‐neighbor}
\acrodef{KDE-1NN}[KDE-1NN]{kernel density estimation with 1-nearest-neighbor}
\acrodef{GNSS}[GNSS]{global navigation satellite systems}

\theoremstyle{definition}

\usepackage{algorithm}
\usepackage{algpseudocode}
\algnewcommand{\LeftComment}[1]{\Statex \(\triangleright\) \textit{#1}}

\usepackage{soul,color}

\usepackage{ragged2e}
\usepackage{blindtext}

\newcommand{\blue}[1] {{\color{black}{#1}}}
\newcommand{\green}[1] {{\color{black}{#1}}}

\newcommand{\blueR}[1] {{\color{black}{#1}}}

\DeclareMathOperator*{\argmin}{argmin}

\begin{document}

\title{Improving Outdoor Multi-cell Fingerprinting-based Positioning via Mobile Data Augmentation} %

\author{Tony~Chahoud,~\IEEEmembership{Member,~IEEE},~Lorenzo~Mario~Amorosa,~\IEEEmembership{Member,~IEEE},\\
Riccardo~Marini,~\IEEEmembership{Member,~IEEE},~and~Luca~De~Nardis,~\IEEEmembership{Member,~IEEE}

\thanks{
\indent T. Chahoud and R. Marini are with CNIT/WiLab - National Wireless Communication Laboratory, Bologna, Italy. E-mail: \{tony.chahoud, riccardo.marini\}@wilab.cnit.it \\
\indent L.M. Amorosa is with the Department of Electrical, Electronic and Information Engineering (DEI), ``Guglielmo Marconi", University of Bologna \& CNIT/WiLab - National Wireless Communication Laboratory, Bologna, Italy. E-mail: lorenzomario.amorosa@unibo.it \textit{(Corresponding Author)}\\
\indent L. De Nardis is with the Department of Information Engineering, Electronics and Telecommunications, Sapienza University of Rome, Rome, Italy \& CNIT/WiLab - National Wireless Communication Laboratory, Bologna, Italy. E-mail: luca.denardis@uniroma1.it
}
}

\markboth{}{}

\maketitle

\begin{abstract}
Accurate outdoor positioning in cellular networks is hindered by sparse, heterogeneous measurement collections and the high cost of exhaustive site surveys. This paper introduces a lightweight, modular mobile data augmentation framework designed to enhance multi-cell fingerprinting-based positioning using operator-collected minimization of drive test (MDT) records. The proposed approach decouples spatial and radio-feature synthesis: kernel density estimation (KDE) models the empirical spatial distribution to generate geographically coherent synthetic locations, while a \blueR{1-nearest-neighbor (1NN)}-based block produces augmented per-cell radio fingerprints. The architecture is intentionally training-free, interpretable, and suitable for distributed or on-premise operator deployments, supporting privacy-aware workflows. 
We both validate each augmentation module independently and assess its end-to-end impact on fingerprinting-based positioning using a real-world MDT dataset provided by an Italian mobile network operator across diverse urban and peri-urban scenarios.
Results show that the proposed KDE-\blueR{1NN} augmentation consistently improves positioning performance with respect to state-of-the-art approaches, reducing the median positioning error by up to 30\% \blueR{(specifically achieved at an augmentation rate of $\times 30$ in the highly sparse highway scenario)}. We also observe region-dependent saturation effects, which emerge most rapidly in scenarios with high user density where the information gain from additional synthetic samples quickly diminishes. Overall, the framework offers a practical, low-complexity path to enhance operator positioning services using existing mobile data traces.
\end{abstract}

\IEEEpeerreviewmaketitle

\section{Introduction}
\label{sec:introduction}

\IEEEPARstart{R}{ecent} years have seen \blue{a growing demand for accurate and reliable positioning services in dense urban areas and indoor environments \cite{yang2024positioning}.
While \blueR{\ac{GNSS}} like the \ac{GPS} remain the gold standard for global positioning, their utility in operational mobile networks is frequently constrained by practical factors, as they remain power-intensive components~\cite{karki2020power}. Consequently, mobile operating systems and user policies often disable or duty-cycle satellite tracking to conserve battery life.}
In these scenarios, multicell fingerprint-based positioning has emerged as a promising approach due to its robustness in non-line-of-sight conditions. However, statistical insufficiency of training data is one of the core limitations of fingerprinting. It manifests in coarse positioning granularity, especially in cell‐dense regions where small geography can exhibit rapid changes in received signal levels, affecting positioning algorithms' performance. Moreover, collecting exhaustive user traces raises privacy concerns, as these measurements may encode sensitive movement patterns \cite{gumble}. These challenges motivate the generation of synthetic mobile measurement data that can augment real traces, filling spatial gaps in the fingerprint database while mitigating the need for additional data collection campaigns.

To improve positioning performance in cellular networks, we introduce a mobile data augmentation framework that synthesizes large volumes of realistic radio measurements. 
Our proposed computationally-light mobile data generator combines \ac{KDE} for estimating spatial components with non‐parametric regression leveraging \blueR{\ac{1NN}} for multi‐cell \ac{RSRP} features generation. 
This methodology aims to produce dense, geographically coherent synthetic mobile data that mimics the statistics of real-world measurements.
Although this framework is designed and tested end-to-end for fingerprinting-based positioning, the augmented measurements it generates may also benefit other tasks, including radio-environment map reconstruction, training and validation of learning-based models for coverage/capacity estimation and link-quality prediction, data-efficient handover and beam-management policy learning, and anomaly/fault detection in network monitoring \cite{ayanoglu2022machine}.
\blueR{Crucially, while prevailing trends in data augmentation heavily favor highly parameterized deep generative architectures, we reveal a critical limitation within the mobile computing domain: these complex models structurally fail when applied to sparse and context-agnostic outdoor radio measurements. Identifying this fundamental bottleneck directly motivates our architectural design. We demonstrate that in such constrained operational regimes, a non-parametric approach with strong local inductive biases is not only a lightweight alternative, but a necessity to accurately synthesize irregular spatial variations without hallucinating non-existent radio features.}

In this work, by leveraging real-world, large-scale \ac{MDT} data provided by an Italian \ac{MNO}, we systematically investigate the positioning performance gains introduced by our proposed synthetic data augmentation pipeline, quantifying improvements in positioning accuracy and robustness across diverse urban scenarios.

\subsection{Contributions and Paper Organization}

This work proposes a novel data augmentation framework specifically designed to enhance fingerprinting-based positioning accuracy in outdoor cellular environments. Our key contributions are summarized as follows:

\begin{itemize}
    \item \blue{Unlike previous studies that focused primarily on synthetic data augmentation to improve positioning in controlled or indoor scenarios, this work addresses the practical challenges inherent to operational outdoor environments. We validate the proposed framework on real \ac{MDT} data provided by an Italian \ac{MNO} in four large-scale settings, demonstrating its capability to handle sparse spatial sampling, heterogeneous and partial \ac{PCI} visibility, measurement noise, and strong temporal/spatial variability typical of user mobility and wide-area propagation effects.}
	
    \item The proposed framework introduces a hybrid and modular model that decouples the generation of user locations and radio features, aiming for a lightweight and interpretable architecture. The user spatial distribution is modeled using \ac{KDE}, while \ac{PCI}-specific radio fingerprints are generated through \blueR{\ac{1NN}, identified as the best-performing neighbor-count configuration among those evaluated}.
    
	\item We empirically validate each module of the augmentation pipeline in isolation. For the \emph{spatial} augmentation stage, we compare \ac{KDE} against parametric and learned alternatives (\ac{GMM}, \ac{GAN}, and \ac{NF}) using a multivariate two-sample \ac{KS} test to quantify distributional agreement with real MDT locations. {For the \emph{radio} augmentation stage, we benchmark the proposed \blueR{\ac{1NN}} approach against \blueR{alternative nearest neighbors approaches,} \ac{RF}, \ac{GPR} with both \ac{SE} and \ac{RQ} kernels, \ac{MLP}, \ac{CGAN}, \ac{ReFlow} (diffusion model), and \ac{NF}. We assess prediction quality in terms of \ac{MAE} on per-\ac{PCI} \ac{RSRP} values.}
    
	\item \blue{We assess the end-to-end impact of augmentation on multi-cell fingerprinting-based positioning by comparing the proposed \ac{KDE}-\blueR{\ac{1NN}} pipeline against representative two-stage baselines (\ac{KDE}-\ac{RF}, \ac{KDE}-\ac{GPR}, \blueR{KDE-2NN, KDE-3NN, KDE-4NN, and KDE-5NN}) and advanced deep learning architectures (\ac{KDE}-\ac{MLP}, \ac{KDE}-\ac{CGAN}, \ac{KDE}-\ac{ReFlow}, and \ac{KDE}-\ac{NF}) by means of average positioning errors over multiple tests.} \blueR{We demonstrate that in context-agnostic outdoor scenarios, non-parametric methods with strong local inductive biases fundamentally outperform state-of-the-art deep generative models.}

	\item We quantify how positioning performance scales with the amount of synthetic data. By sweeping augmentation rates and performing pairwise statistical comparisons over positioning errors obtained by our architecture, we identify region-dependent saturation points: beyond a scenario-specific augmentation level, additional synthetic samples yield diminishing returns and, in some cases, may not improve.
    
	\item To facilitate \blueR{replicability} and support further research in the field, we make the source code of the proposed framework publicly available\footnote{\blueR{Real-world MDT traces provided by the \ac{MNO} are confidential and cannot be publicly disclosed due to user privacy. Therefore, the dataset included in the repository is anonymized and features artificial coordinates.}}. The repository can be accessed at: \url{https://github.com/wilabcnit/Mobile-Data-Augmentation-Positioning}.
\end{itemize}
The remainder of the paper is structured as follows.
Sec.~\ref{sec:background} provides the background on fingerprinting and mobile data collection.
Sec.~\ref{sec:related_work} reviews the related work.
Sec.~\ref{sec:data_augm} introduces the proposed mobile data augmentation pipeline, describing the modular components for the spatial and radio feature generation.
Sec.~\ref{sec:positioning-scenarios} discusses both the multi-cell fingerprinting-based positioning problem and introduces the four real-world \ac{MDT}-based reference scenarios.
Sec.~\ref{sec:results} presents the numerical results in terms of (i) independent performance of each stage of the modular framework, (ii) performance benefits of augmentation to fingerprinting-based positioning in the considered scenarios, and (iii) analysis on performance scaling with respect to the amount of synthetic data generated.
Finally, Sec.~\ref{sec:conclusion} concludes the paper.

\section{Background}
\label{sec:background}

\subsection{Fingerprinting in Cellular Networks}
Fingerprint-based positioning leverages existing cellular infrastructure and has become a standard for robust positioning in dense urban and indoor environments \cite{fp1, fp2}. The \ac{3GPP} incorporated fingerprinting-based positioning into LTE networks under Release 12~\cite{3gpp36809}. This approach estimates a device's position by comparing its observed radio measurements, such as \ac{RSRP}, to a pre-constructed database of location-tagged fingerprints. By leveraging existing cellular infrastructure, fingerprinting provides robust performance where signal blockage and multipath propagation often degrade the accuracy of conventional techniques. 
However, achieving high positioning accuracy through multi-cell fingerprinting requires large volumes of spatially and radio‐feature‐rich data in order to capture fine‐grained variations across coverage areas; in practice, collecting such dense measurement sets is both costly and time‐consuming for \acp{MNO}~\cite{deNardis2023}, which poses a significant barrier to widespread deployment.

\subsection{Minimization of Drive Tests (MDT)}
To enable data-driven optimization, we leverage the \ac{MDT} framework introduced in \ac{3GPP} Release 10 \blue{\cite{3gpp_mdt_release_10}}. In \ac{MDT}, user equipments periodically report field measurements, including \ac{RSRP}, into the network's \ac{OM} system in order to facilitate failure forecasting, troubleshooting, and network optimization \blue{\cite{skocaj2022cellular, qureshi2020enhanced}}. 
\blue{While this framework provides a rich, cost-efficient alternative to traditional drive testing, it is subject to the operational constraints of commercial user devices. Notably, in the commercial \ac{MDT} dataset analyzed in this work, valid \ac{GPS} coordinates are present in fewer than 4\% of the reported samples, as devices frequently disable satellite positioning to conserve power.} This sparsity necessitates the development of augmentation strategies that can effectively utilize the limited available geolocated data.

\section{Related Work}
\label{sec:related_work}

Data augmentation has emerged as a powerful technique across several domains, including image classification~\cite{krizhevsky2012imagenet,lecun1998gradient} and audio recognition~\cite{schluter2015exploring}, to mitigate data scarcity, enhance generalization, and improve model robustness. In the context of wireless positioning, these challenges are even more pronounced due to the high cost and manual effort involved in collecting ground-truth fingerprints in real-world environments, motivating the development of augmentation strategies that can synthetically expand datasets while preserving the spatial and statistical characteristics of the underlying radio environment.
While established geo-statistical estimators like Kriging explicitly model spatial correlation to produce geo-tagged features at unmeasured locations \cite{kriging}, these approaches often have high sensitivity to model assumptions, rely on the assumption of data stationarity and isotropy, and become computationally intensive for very large datasets.

%
Recent \blueR{surveys~\cite{wifi_survey, fathalizadeh2025survey} have} documented various augmentation strategies developed for WiFi-based indoor fingerprinting, such as spatial interpolation and signal transformations by leveraging generative models. \blueR{Recent advancements further demonstrate the effectiveness of graph attention networks~\cite{gufran2025gate} and light-based fingerprint synthesis~\cite{lee2025generating} in handling data sparsity for complex indoor layouts. Additionally, recent studies emphasize the critical relevance of hyperparameter optimization in similarity-based and hybrid topological algorithms for dynamic environments~\cite{shir2025topology}.} However, these approaches are proven to be effective in dense indoor settings using short-range signal features, while their efficacy in outdoor cellular scenarios, which involve sparse, noisy, and heterogeneous measurements, is not thoroughly investigated yet. 
\blue{Transferring these methods to the outdoor domain is non-trivial due to distinct operational characteristics: unlike static indoor layouts, wide-area cellular networks exhibit significant variability in macro-cell density and coverage geometry, alongside practical anomalies such as incomplete \ac{PCI} visibility and unpredictable user mobility patterns \cite{s20020427}.}

Several positioning approaches, which typically rely on signal propagation models~\cite{mazuelas2009robust,yang2009indoor} or fingerprinting systems supported by specialized hardware deployments~\cite{ji2006ariadne,mirowski2012depth}, can yield accurate results in controlled environments, but they are difficult to scale in large outdoor areas where exhaustive site surveys are impractical and radio propagation is highly variable in space and time. 
\blue{Furthermore, operational outdoor networks are subject to continuous temporal drift driven by environmental non-stationarity (e.g., seasonal changes), traffic-dependent interference, and long-term network evolution, all of which degrade the longevity and reliability of static fingerprint databases \cite{Gong2023}.} 
As such, there remains a pressing need for augmentation methods tailored specifically to outdoor cellular positioning challenges.
Building on this indoor-focused literature, \ac{GPR} has been explored for augmenting WiFi fingerprints in~\cite{sun2018augmentation} to improve positioning performance. 
Although initially designed for indoor settings, we nonetheless treat \ac{GPR} as a meaningful baseline for radio-feature augmentation and include it in our experimental comparisons to quantify its strengths and limitations on outdoor data. 

It is worth noting that radio features can also be predicted using deterministic approaches, such as \ac{RT}~\cite{hoydis2023sionna, 10419169}, or empirical propagation models~\cite{yapar2024overview}. While RT can provide accurate channel estimation, it requires precise 3D environmental models (digital twins) and detailed infrastructure metadata (e.g., antenna patterns, material properties), which are computationally expensive to process and rarely available for large-scale operational networks. Similarly, standard empirical models often lack the granularity to capture local shadowing effects essential for fingerprinting. Furthermore, both approaches typically model the physical channel rather than the reported measurement data, without accounting for device heterogeneity, body-loss, and measurement noise present in real \ac{MDT} traces. In contrast, data-driven approaches allow for feature augmentation that implicitly learns these complex statistical distributions directly from the data.

In a recent study, the GUMBLE framework~\cite{gumble} has been proposed as a general-purpose generative model for mobile network data using Bayesian conditional inference. While effective in learning conditional relationships in network data, this framework
is not tailored to generate synthetic data to improve 
positioning accuracy; instead, it focuses on quantifying the aleatoric and epistemic uncertainty of a synthetically generated mobile dataset.
Other recent generative frameworks have explored conditioning the signal generation process on auxiliary environmental features such as terrain, cell orientation, and transmission characteristics~\cite{sun2022gendt}, while other approaches have leveraged satellite imagery or \ac{GPS} traces to estimate signal quality at unmeasured locations~\cite{thrane2018drive}.
However, such contextual inputs are rarely available in practice.
Recent literature supports the view that complex generative models may not be optimal for all scenarios. For instance, the survey in \cite{qureshi2023toward} indicates that for low-dimensional, non-smooth radio data where network geometry is unknown, interpolation-based approaches can offer superior robustness compared to deep learning methods. 
\blueR{While existing fingerprinting augmentation techniques perform well under assumptions of dense spatial sampling, stationary environments, or the availability of auxiliary metadata (e.g., 3D building maps, antenna geometries), operational MDT data violates these assumptions. The fundamental challenge addressed in this work is generating statistically faithful augmented radio maps relying only on highly sparse GPS tags and RSRP measurements without any environmental context.}
Building on this insight, and in contrast with the existing literature, the focus of our work is to develop a data-driven, context-agnostic augmentation framework specifically tailored for positioning in large-scale outdoor settings. Our proposed method requires only geo-tagged datasets without auxiliary environmental metadata, and enables the generation of synthetic user locations and radio features in a manner that reflects the true characteristics of real-world mobile datasets to enhance positioning accuracy, particularly in urban regions where data scarcity and signal irregularities are most pronounced.

\section{Mobile Data Augmentation}
\label{sec:data_augm}
\begin{figure*}[ht]
    \centering
    \includegraphics[width=\textwidth]{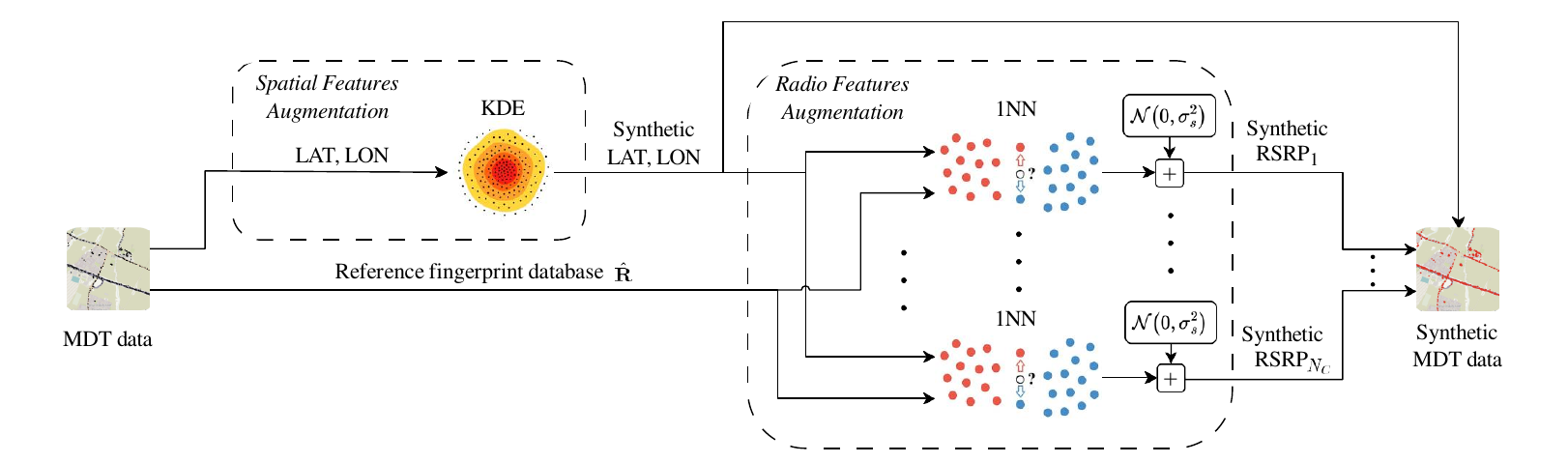}
    \caption{Mobile data augmentation architecture, comprising \emph{Spatial Features Augmentation} and \emph{Radio Features Augmentation} stages. 
    The augmented dataset is ready to use for fingerprinting-based positioning algorithms.}
    \label{fig:mobile_data_augmentation}
\end{figure*}

The goal of the proposed mobile data augmentation architecture is to synthesize large volumes of geographically coherent, multi-cell radio features that (i) fill spatial gaps in operator-collected traces and (ii) preserve the statistical characteristics of real measurements, so that the augmented data can be used directly by fingerprinting-based positioning algorithms. The architecture is intentionally computationally light and comprises two modular stages: \emph{Spatial Features Augmentation} and \emph{Radio Features Augmentation}. Fingerprints built from geo-referenced per-cell power measurements have been widely used for positioning across data sources, ranging LTE-\ac{RSRP} collected in field trials and drive tests~\cite{RSRPfingerprint, fp1} to scanner-based datasets~\cite{deNardis2023}. While broadly applicable across mobile-network data sources, we focus on \ac{MDT} as a representative case due to its \ac{3GPP} compliance and widespread adoption by \acp{MNO}~\cite{3gpp_mdt_release_10}. However, it is important to note that our approach is not specific to MDT and can be applied to any geo-tagged dataset with per-cell power measurements, such as those collected through mobile network scanners, drive tests, or other crowdsourced data campaigns.

Starting from a set of geo-tagged \ac{MDT} records collected by the \ac{MNO}, the proposed pipeline proceeds in two stages. 
First, \ac{KDE} is used to estimate the empirical spatial distribution of measurements and to draw synthetic sample locations. 
Second, for each synthetic location, a multi-cell \ac{RSRP} vector is predicted via a \blueR{\ac{1NN}} operating on the original MDT fingerprints.
This produces augmented \ac{MDT} data that can be integrated into fingerprinting-based positioning algorithms.
The overall flow is illustrated in Fig.~\ref{fig:mobile_data_augmentation}. MDT data structure is discussed in Sec.~\ref{subsec:mdtData}, while the two-stage architecture details are given in Sec.~\ref{subsec:spatialAugm} and Sec.~\ref{subsec:radioAugm}.

\begin{table}[!t]
\centering
\renewcommand{\arraystretch}{1.5}
\ttfamily
\rowcolors{2}{gray!5}{white} 
\resizebox{\columnwidth}{!}{%
\begin{tabular}{c c c c c c c}
\hline
\textbf{Longitude} & \textbf{Latitude} & \textbf{RSRP\_PCI\_1} & \textbf{RSRP\_PCI\_2} & \textbf{RSRP\_PCI\_3}  & $\dots$ \\ \hline
11.3456 & 44.4945 & -87 dBm & -95 dBm & -- &$\dots$ \\
11.3460 & 44.4951 & -90 dBm & -92 dBm & -105 dBm & $\dots$ \\
11.3465 & 44.4958 & -85 dBm & -- & -102 dBm  & $\dots$ \\
11.3470 & 44.4963 & -88 dBm & -96 dBm & -113 dBm & $\dots$ \\ 
$\vdots$ & $\vdots$ & $\vdots$ & $\vdots$ & $\vdots$ & $\ddots$ \\ \hline
\end{tabular}%
}
\caption{Representative sample of the MDT dataset.}
\label{tab:mdt}
\end{table}

\subsection{MDT Data}
\label{subsec:mdtData}
This study leverages a real-world dataset collected in \ac{MDT} format by a commercial Italian \ac{MNO}. 
The core radio feature in \ac{MDT} data utilized for positioning is the \ac{RSRP}, which quantifies the received power of the \ac{LTE} \ac{CRS} as measured at the \ac{UE}. \ac{RSRP} is particularly suitable for positioning applications due to its insensitivity to co-channel interference and average network load. This robustness arises from the \acp{UE} ability to isolate and correlate the \acp{CRS}, thereby extracting \ac{PCI}-specific signal strength even in complex radio conditions.

Formally, the \ac{RSRP} at a given location is defined as the average power measured over all \acp{RE} that carry \acp{CRS} across the system bandwidth \cite{3GPP_36.214}:
\begin{equation}
    \text{RSRP} = \frac{1}{N} \sum_{i=1}^{N} \sum_{k=1}^{14} P_{\text{RE},ik},
    \label{eq:rsrp}
\end{equation}
where \( N \) is the number of subcarriers and \( P_{\text{RE},ik} \) denotes the received power for subcarrier \( i \) and \ac{OFDM} symbol \( k \)\green{, with \( k \) ranging over the 14 \ac{OFDM} symbols of a 1\,ms \ac{TTI} under the LTE normal cyclic prefix}.
The value of \( P_{\text{RE},ik} \) is determined as:
\begin{equation}
    P_{\text{RE},ik} =
    \begin{cases}
        P_{\text{RE},i} & \text{if the } k\text{-th OFDM symbol carries CRS}, \\
        0 & \text{otherwise}.
    \end{cases}
\end{equation}
\ac{RSRP} serves as a critical input to multiple \ac{RRM} procedures, including mobility control, cell reselection, and handover decisions. As a spatially informative and locally averaged indicator, it provides a stable reference for positioning tasks, particularly when collected across multiple cells.\\

In the dataset, \ac{RSRP} measurements are reported not only for the serving \ac{PCI}, but also for all detectable neighboring \acp{PCI} within the reception range of the corresponding \ac{UE}. This structure, exemplified in Table ~\ref{tab:mdt}, provides a direct representation of the measurement format.
Each entry is associated with a geographical position expressed as a latitude-longitude pair and \ac{RSRP} measurements for each distinct \ac{PCI} observed within the considered region. 
For positions where a given \ac{PCI} is not detected, the corresponding measurement is absent, resulting in a sparse data structure that reflects realistic coverage conditions.

\subsection{Spatial Features Augmentation via Density Estimation}
\label{subsec:spatialAugm}

Tabular data generation, including spatial features augmentation, can be addressed using a wide range of approaches \cite{sdv}. State-of-the-art methods include \acp{GAN} \cite{razghandi2022vaeicc, francesco2018assembling}, \acp{NF} \cite{Rezende2015nf}, \acp{VAE} \cite{ctgan-tvae}, and \acp{GMM} \cite{li2018high}. However, the spatial distribution of \ac{MDT} data presents specific challenges for these models due to its inherent complexity, irregularity, and strong dependence on the topography of the area under investigation.

To this end, we propose modeling user sample generation through
a bi-variate density estimation approach over the latitude and longitude components of \ac{MDT} data. 
Specifically, we can set the density estimation problem as a maximum log-likelihood problem:
\begin{equation}\label{eq:ml_estimation}
\theta^* = \argmin_{\theta}\ \mathbb{E}_{\mathbf{p}\sim P(\mathbf{p})}\bigl[-\log f(\mathbf{p}\mid \theta)\bigr]\,,
\end{equation}
where $\mathbf{p} \in \mathbb{R}^2$ denotes the latitude and longitude of a data point sampled from the true distribution $P(\mathbf{p})$, $\theta$ is the model parameterization, and $f(\mathbf{p}\mid \theta)$ is the modeled density.
In practice, we approximate the expectation via \ac{ERM} over a training set $\mathcal{D}=\{\mathbf{p}_i\}_{i=1}^m$, yielding the empirical loss
\begin{equation}\label{eq:empirical_loss}
\mathcal{L}(\theta) = -\sum_{i=1}^m \log f(\mathbf{p}_i\mid \theta)\,,
\end{equation}
and optimize
$\theta^* = \argmin_{\theta} \mathcal{L}(\theta)$.

Specifically, we adopt \ac{KDE}, a non-parametric technique, motivated by its simplicity, interpretability, and suitability for modeling the spatial distribution of users. 
\blue{Unlike black-box generative models, \ac{KDE} offers a transparent decision mechanism where the estimated density is directly traceable to the proximity of specific training examples.} 
\green{Furthermore, \ac{KDE} operates only on geo-referenced coordinates without requiring contextual information, such as base-station positions, antenna orientation, or other network-side geometric information.} 
In Sec.~\ref{sec:abl_spatial}, we further validate this choice by comparing performance using the multivariate \ac{KS} test against parametric baselines such as \acp{GAN}, \acp{NF}, and \acp{GMM}.
Specifically, \ac{KDE} estimates the underlying \ac{PDF} by placing a kernel $K(\mathbf{p} - \mathbf{p}_i, h)$ centered at each training point $\mathbf{p}_i \in \mathcal{D}$, with $h$ representing the kernel bandwidth. This yields an interpretable \ac{PDF} estimate $f(\mathbf{p}, \mathcal{D}, h): \mathbb{R}^2 \rightarrow \mathbb{R}^+$, constructed directly from the empirical data, capturing multi-modal structures inherent in the distribution \cite{kde}. Formally, the \ac{KDE} is defined as
\begin{equation}\label{eq_kde}
f(\mathbf{p}, \mathcal{D}, h) = \frac{1}{m}\sum\limits_{i=1}^{m} K(\mathbf{p} - \mathbf{p}_i, h)\,,
\end{equation}
where the kernel is given by the Gaussian function:
\begin{equation}
K(\mathbf{p} - \mathbf{p}_i, h) = \frac{1}{h\sqrt{2\pi}} \exp\left(-\frac{(\mathbf{p} - \mathbf{p}_i)^2}{h^2}\right)\,.
\end{equation}
Here, $\mathbf{p}$ is the evaluation point, and the model parameters $\theta$ correspond to the training data points $\mathcal{D}$.

We select the Gaussian kernel based on domain-specific considerations. MDT location data are typically obtained via GPS at the user side, and the associated measurement noise is typically modeled as Gaussian. Therefore, using a Gaussian kernel with a tunable bandwidth aligns naturally with the assumed uncertainty in user positioning. 
\blue{This reinforces the model's interpretability: the predicted density at any point $\mathbf{p}$ is not an abstract score, but a cumulative measure of how many training users were physically located within the noise radius $h$ of that point.} 
Intuitively, \ac{KDE} assumes that regions with higher sample density are more likely to generate future observations. Since \ac{KDE} is an unsupervised method, the bandwidth $h$ is treated as a hyperparameter and tuned via \ac{ERM}. Specifically, we minimize the negative log-likelihood on a validation subset $\mathcal{D}' = \{\mathbf{p}_i'\}_{i=1}^{m'}$, that is
\begin{equation}\label{eq_ll_bandwith}
\argmin_h -\sum_{i=1}^{m'} \log f(\mathbf{p}_i', \mathcal{D}, h)\,.
\end{equation}

\subsection{Radio Features Augmentation via Non-Parametric Regression}
\label{subsec:radioAugm}

Radio features augmentation can be addressed with a range of approaches; representative examples include neural networks \cite{gumble}, \acp{RF} \cite{rf_localization}, and \acp{GPR} \cite{sun2018augmentation}, which have been used for related tasks in literature.
In this work, we adopt \blueR{\ac{1NN}}, which is an interpretable non-parametric procedure that transfers measured per-PCI radio features from nearby reference locations.
\blue{In Sec.~\ref{sec:abl_radio}, we further validate this choice by comparing performance using \ac{MAE} on radio feature prediction against machine learning approaches (\ac{RF} and \ac{GPR}) and deep learning approaches (\ac{MLP}, \ac{CGAN}, \ac{ReFlow}, and \ac{NF}).}
\blueR{Furthermore, we evaluate the impact of varying the number of nearest neighbors in the set $\{1, 2, 3, 4, 5\}$, selecting 1NN based on validation-set performance.}

Let $N_c$ denote the total number of cells in the region of interest. 
We assume access to a reference fingerprint database
\begin{equation}
  \green{\mathbf R}
  = \bigl\{(\green{\mathbf r_j},\;\green{\mathbf p_j})\bigr\}_{j=1}^m
  \quad\text{with}\quad
  \green{\mathbf r_j} \in \mathbb R^{N_c},\;\green{\mathbf p_j}\in\mathbb R^2,
\end{equation}
where each $\green{\mathbf r_j}$ is a measurement vector recorded at the known location $\green{\mathbf p_j}$.
Given a query spatial location $\mathbf p\in\mathbb R^2$ (either sampled from a test set, sampled from the KDE in Sec.~\ref{subsec:spatialAugm}, or otherwise selected), our goal is to synthesize a realistic multi-cell radio fingerprint
\(\green{\hat{\mathbf r}}(\mathbf p) = [ \green{\hat r_{1}}(\mathbf p),\dots, \green{\hat r_{N_c}}(\mathbf p)]^\top\)
consisting of per-PCI \ac{RSRP} measurements that are consistent with the empirical database
\(\green{\mathbf R}\).

We use a spatial nearest-neighbor rule to transfer radio measurements from the reference database to the query point. We denote the index of the training location closest in Euclidean distance to the query point $\mathbf p$ as
\begin{equation}
    j^*(\mathbf p) \;=\; \argmin_{j=1,\dots,m} \bigl\lVert \mathbf p - \green{\mathbf p_j} \bigr\rVert_2\,,
\end{equation}
\blueR{where $\mathbf p$ denotes local planar coordinates (in meters), obtained via haversine-based projection from a fixed reference point in each region, and $\bigl\lVert \cdot \bigr\rVert_2$ denotes the Euclidean distance.} 
The nearest training fingerprint $\green{\mathbf r_{j^*}}$ provides the baseline per-PCI measurements to be assigned to $\mathbf p$.
To model slow, large-scale fluctuations caused by obstacles, buildings and other shadowing phenomena, we perturb the assigned baseline \ac{RSRP} values using a log-normal (additive Gaussian in the dB scale) shadowing term, which is the standard parametrization in the propagation literature \cite{3gpp_38_901}. Concretely, for each PCI $c$, we draw an independent shadowing sample
\begin{equation}
    s_c \;\sim\; \mathcal{N}\bigl(0,\sigma_s^2\bigr),
\end{equation}
and define the augmented (continuous) measurement
\begin{equation}\label{eq:augmented_rsrp}
     \green{\hat r_{c}(\mathbf p)} =  \green{r_{j^*,c}} + s_c\,.
\end{equation}
Notably, if a PCI is not detected in the $j$-th training location, we simply leave it unassigned, i.e., no measurement is generated for that PCI at $\mathbf p$.

\subsection{Analysis and Benefits}
\label{subsec:advantages}

The proposed mobile data augmentation pipeline is intentionally simple and modular: a centralized \emph{Spatial Features Augmentation} stage (KDE-based) produces geographically coherent locations, while a lightweight \emph{Radio Features Augmentation} stage (\blueR{1NN}-based) transfers and perturbs empirical per-PCI \ac{RSRP} fingerprints. This separation of concerns is the starting point of several practical advantages that make the architecture attractive for operator-grade deployment.

\begin{enumerate}
    \item The architecture naturally supports a \textit{hybrid deployment model}. The spatial KDE, requiring a global view of geo-tagged reports to accurately capture multi-modal spatial densities, is best executed centrally (e.g., within the operator back-end or an on-premise data-hub). Conversely, the radio-transfer \blueR{1NN}-based module can be executed in a distributed manner (e.g., at the network edge, in regional data-centers, or even on-site at measurement collection points) because it only needs access to a local fingerprint catalogue or indexed subset of 
    \(\green{\mathbf R}\).
    This division reduces communication overhead: only the spatial model or sampled synthetic locations must be exchanged, while radio fingerprints can be generated locally when required.
    \item The pipeline is essentially \emph{training-free} for radio-feature synthesis. KDE sampling and the nearest-neighbor algorithm avoid iterative, parameter-intensive training loops typical of RF, \ac{GPR}, or deep models. The practical consequences are significant: lower computational and engineering costs, faster time-to-deployment, and elimination of long retraining cycles whenever new MDT records become available. Because no complex model parameters must be learned and propagated, newly collected data can be appended and used immediately to refresh the empirical database or to re-sample the KDE, providing a seamless data-integration path that directly benefits both augmentation and downstream positioning.
    \item The approach yields high \textit{interpretability} and simplifies validation. KDE bandwidth, kernel choice, and the \blueR{1NN} algorithm are explicit, easy-to-inspect hyperparameters whose effects on the generated distribution are intuitive. This transparency facilitates validation and supports operator certification workflows, where explainability and repeatable test procedures are often mandatory. In practice, this means operators can reliably audit the augmentation process, identify failure modes, and enforce quality gates before synthetic data are admitted into production positioning databases.
    \item The design supports \textit{privacy- and data-minimization} strategies. Since the spatial model can be represented by a compact density parameterization (or by synthetic locations sampled from it), operators can avoid sharing raw, user-level MDT traces across different sites. Moreover, in distributed deployments, the detailed per-PCI measurements can remain in the data domain where they were collected, reducing privacy exposure and simplifying compliance with data-protection requirements.
    \item The modular architecture is \textit{extensible}: the spatial features augmentation and the radio features augmentation components can be replaced or augmented with other predictors when computational resources, data volume, or fidelity requirements justify it. Because modules expose clear interfaces (i.e., the input/output scheme at each component), alternative models can be integrated or benchmarked without redesigning the whole generator or the downstream positioning pipeline.
    \item From an operational standpoint, the combination of \textit{low runtime cost} and \textit{immediate usability of newly collected data} makes the KDE-\blueR{1NN} solution particularly suitable for on-premise or edge deployments where compute resources are constrained and timeliness is important.
\end{enumerate}
Taken together, these properties explain why a lightweight, statistically faithful augmentation pipeline can be a practical and robust choice for improving fingerprinting-based positioning in real-world MDT-driven scenarios.

We note that the spatial- and radio-feature augmentation modules admit independent design choices and hyperparameters. For example, the \ac{KDE}-based spatial stage can employ different kernel families and bandwidth-selection rules, while the radio stage may use alternative nearest-neighbor schemes (e.g., distance-weighted averaging, or diverse distance metrics) as well as varying shadowing and noise models. A systematic exploration of these alternatives and their impact on downstream positioning performance is deferred to future work.

\section{Fingerprinting-based Positioning and \\Real-World MDT-based Scenarios}
\label{sec:positioning-scenarios}

\subsection{Multi-cell Fingerprinting-based Positioning}
\label{sec:positioning}

There exist several neural network-based approaches to multi-cell fingerprinting-based positioning, including those in \cite{positioning_csi, positioning_mt}. Other works, including  \cite{fp1, deNardis2023,deNardis2025}, rely on non-parametric predictors such as \ac{wKNN} for determining position information.
In this work, we leverage \ac{wKNN} as an illustrative example of a positioning algorithm that can possibly benefit from mobile data augmentation.

In multi-cell fingerprinting-based positioning, a set of base stations (cells) act as anchors whose received‐signal measurements are used to infer an unknown user location. As introduced in Sec.~\ref{subsec:radioAugm}, 
we denote by $\green{\mathbf R}$ the reference fingerprint database, comprising $m$ radio measurements with respect to $N_c$ anchors and user position information.
For the $i$‑th user (at unknown position $\mathbf p_i\in\mathbb R^2$), we observe a radio‑measurement vector
\begin{equation}
    \mathbf r_i = \bigl[r_{i,1},\,r_{i,2},\,\dots,\,r_{i,N_c}\bigr]^\top
  \;\in\;\mathbb R^{N_c}\,.
\end{equation}
To estimate $\mathbf p_i$ from $\mathbf r_i$, we employ the \ac{wKNN} algorithm. 
For any two measurement vectors $\mathbf r_i,\green{\mathbf r_j}\in\mathbb R^{N_c}$,\blueR{ we use a widely adopted distance (e.g., in \cite{wang2019improved}):}
\begin{equation}
 d(\mathbf r_i,\green{\mathbf r_j})
 = \bigl\lvert \mathbf r_i - \green{\mathbf r_j} \bigr\rvert
 = {\sum_{c=1}^{N_c}\;\bigl\lvert\mathbf r_{i,c} - \green{\mathbf r_{j,c}}\bigr\rvert}\,,
\end{equation}
and we perform neighbor selection computing all distances $d\bigl(\mathbf r_i,\green{\mathbf r_j}\bigr)$ for all $j=1,\dots,m$.
We denote the indices of the $K$ reference fingerprints with the smallest distance to $\mathbf r_i$ as
\begin{equation}
   \mathcal N_K(i)
     \;=\;
     \bigl\{\,j_1,\dots,j_K\bigr\}
     \subset \{1,\dots,m\}\,.
\end{equation}
\blueR{In our experimental evaluation, we set the number of neighbors to $K = 3$. A static $K$ strategy is adopted in this work, following the same rationale as~\cite{deNardis2023}, where the impact of $K$ on positioning accuracy was similarly investigated.}
Then, we assign to each neighbor $j\in\mathcal N_K(i)$ a weight
\begin{equation}
    w_j
     = w\bigl(d(\mathbf r_i,\green{\mathbf r_j})\bigr)
     = \frac{1}{\,d(\mathbf r_i,\green{\mathbf r_j})+ \varepsilon\,}\,,
\end{equation}
where $\varepsilon\in \mathbb R^+$ is a small positive constant.
The estimated position $\green{\hat{\mathbf p}_i}$ is the normalized weighted average of the true positions of the $K$ nearest neighbors, that is

\begin{equation}
    \green{\hat{\mathbf p}_i}
     = \frac{\displaystyle
         \sum_{j\in\mathcal N_K(i)} w_j\,\green{\mathbf p_j}
       }{\displaystyle
         \sum_{j\in\mathcal N_K(i)} w_j
       }\,.
\end{equation}  
\blueR{We intentionally adopt \ac{wKNN} as a standard non-parametric baseline for fingerprinting‑based positioning to ensure that any observed performance gains are strictly attributable to the statistical fidelity of the synthetically generated data, rather than being due to algorithmic advancements in the positioning estimator itself.}
In the remainder of this paper, we will assess how augmenting the reference database \(\green{\mathbf R}\) can improve the accuracy of this estimator.

\subsection{MDT-based Reference Scenarios}

In this work, we leverage a real-world \ac{MDT} dataset collected across the city of Bologna (Italy) and provided by an Italian \ac{MNO}. 
In the dataset, each sample corresponds to a user record, including its latitude, longitude, and
\ac{RSRP} measurements \blueR{for the serving \ac{PCI} and at most 8 neighboring \acp{PCI}, the actual number detected varying among the total \acp{PCI} available in each region.}
\blue{To ensure the reliability of these spatial labels subject to \ac{GPS} multipath and drift, we applied a quality control filter based on the positioning uncertainty fields defined in \ac{3GPP} standards. Specifically, we utilized the uncertainty semi-major axis to discard samples with a positioning uncertainty radius exceeding 10 m. This pre-processing step effectively mitigates the impact of location drift, ensuring that the ground-truth coordinates used for training remain robust even in challenging scenarios.} 
The measurements cover four distinct regions, 
namely a dense urban area, mixed urban-rural zone, an airport, and a highway.
To study the impact of augmentation under diverse environmental and operational conditions, the four scenarios are chosen to be heterogeneous.

\begin{figure}[t]
\centering
\begin{minipage}[b]{0.23\textwidth}
    \centering
    \includegraphics[width=\linewidth]{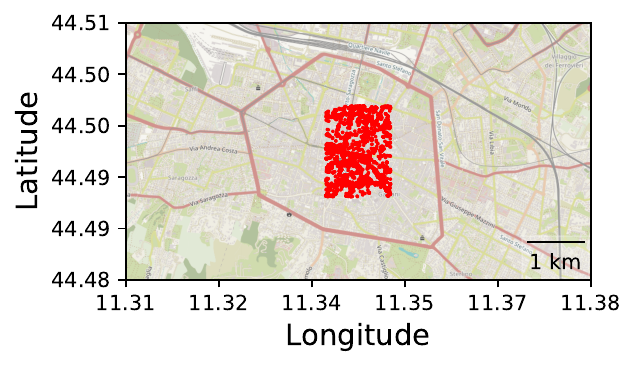}
    \vspace{2pt}
    \small (a) City center
\end{minipage}%
\hspace{5pt}
\begin{minipage}[b]{0.23\textwidth}
    \centering
    \includegraphics[width=\linewidth]{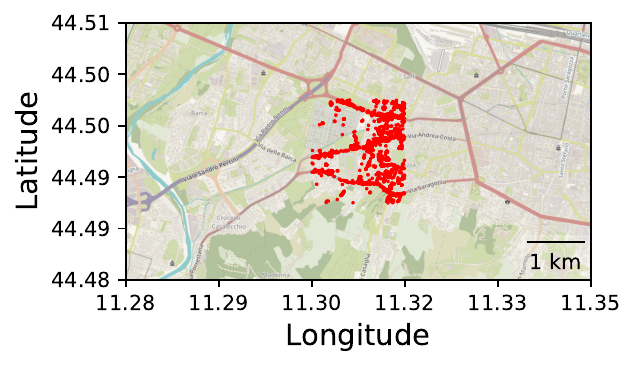}
    \vspace{2pt}
    \small (b) Stadium
\end{minipage}%
\hspace{5pt}
\begin{minipage}[b]{0.23\textwidth}
    \centering
    \includegraphics[width=\linewidth]{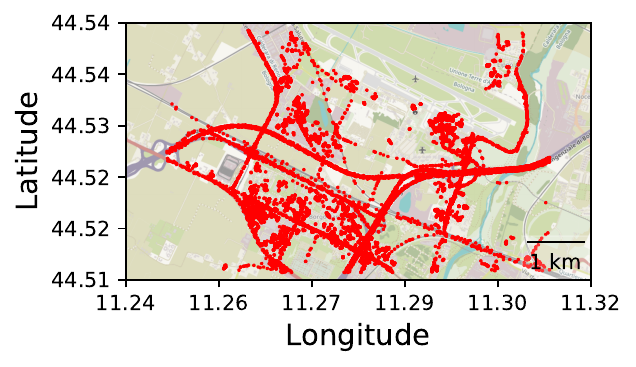}
    \vspace{2pt}
    \small (c) Airport
\end{minipage}%
\hspace{5pt}
\begin{minipage}[b]{0.23\textwidth}
    \centering
    \includegraphics[width=\linewidth]{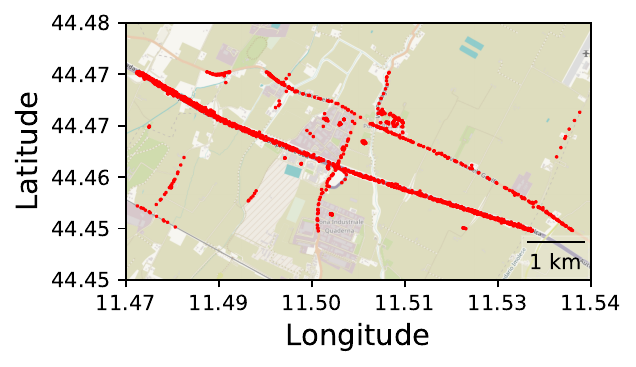}
    \vspace{2pt}
    \small (d) Highway
\end{minipage}
\vspace{6pt}
\caption{The four MDT-based regions in Bologna representing our reference scenarios.}
\label{fig:region_maps}
\end{figure}

These regions vary in spatial density, mobility profiles, and radio coverage characteristics, offering a representative basis for evaluating \ac{RSRP}-based fingerprinting positioning systems. To reflect realistic propagation environments, each area is modeled under predominantly \ac{NLOS} conditions, which are common in urban and suburban deployments. 
\blueR{
Additionally, a stochastic shadow fading perturbation is incorporated into the PCI-wise radio feature modeling process. The shadow-fading variances ($\sigma^2_s$) governing this perturbation were calibrated using the channel models detailed in Section 7.4 of the 3GPP TR 38.901 specification \cite{3gpp_38_901}. Based on this standard, we mapped each of our experimental regions to its corresponding 3GPP propagation scenario under \ac{NLOS} conditions. 
}
Visual representations of these regions are provided in Fig.~\ref{fig:region_maps}. Their quantitative characteristics, including spatial extent, number of \acp{UE}, user density, and 3GPP shadow fading variance \((\sigma_s^2)\), are summarized in Table ~\ref{tab:original_density}.
\label{sec:scenarios}
\begin{table}[t]
\centering
\renewcommand{\arraystretch}{1.2}
\resizebox{\columnwidth}{!}{%
\begin{tabular}{|c|c|c|c|c|}
\hline
\textbf{Region} & \textbf{Area (km$^2$)} & \textbf{Num. of UEs} & \textbf{Density (UE/km$^2$)} & $\boldsymbol{\sigma^2_s}$ \textbf{(dB)} \\
\hline
City center & 2.05  & 6453  & 3147.8 & 7.8 \\
Stadium     & 3.20  & 2705  & 845.3 & 7.8 \\
Airport     & 18.61 & 17698 & 951.0 & 7.8 \\
Highway     & 24.96 & 2920  & 116.9 & 8.0 \\
\hline
\end{tabular}
}
\caption{Quantitative descriptors and shadow fading variance of the four MDT-based regions.}
\label{tab:original_density}
\end{table}
\begin{figure*}[htbp]\centering
\begin{minipage}[b]{0.19\textwidth}
    \centering
    \includegraphics[width=\linewidth,height=2.4cm]{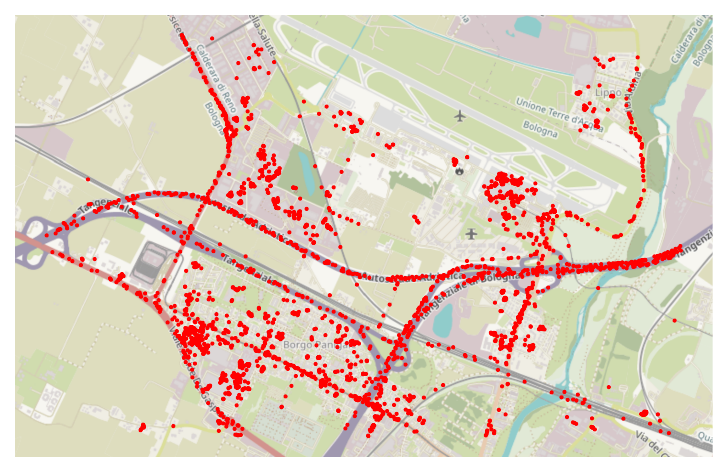}
    \vspace{2pt}
    \small (a) Ground-Truth
\end{minipage}%
\hspace{2pt}
\begin{minipage}[b]{0.19\textwidth}
    \centering
    \includegraphics[width=\linewidth,height=2.4cm]{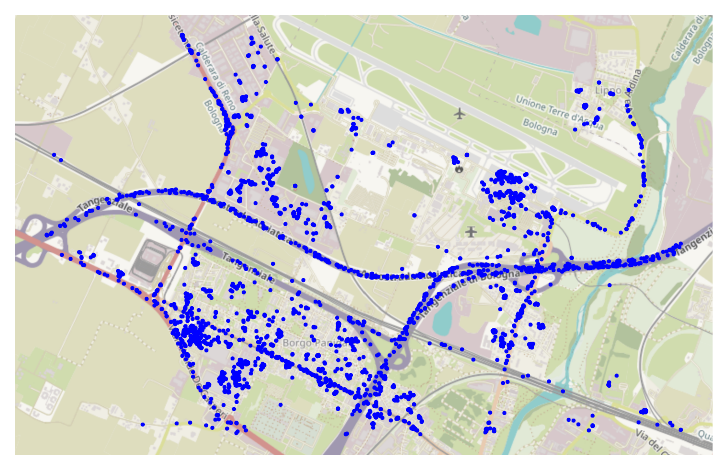}
    \vspace{2pt}
    \small (b) KDE
\end{minipage}%
\hspace{2pt}
\begin{minipage}[b]{0.19\textwidth}
    \centering
    \includegraphics[width=\linewidth,height=2.4cm]{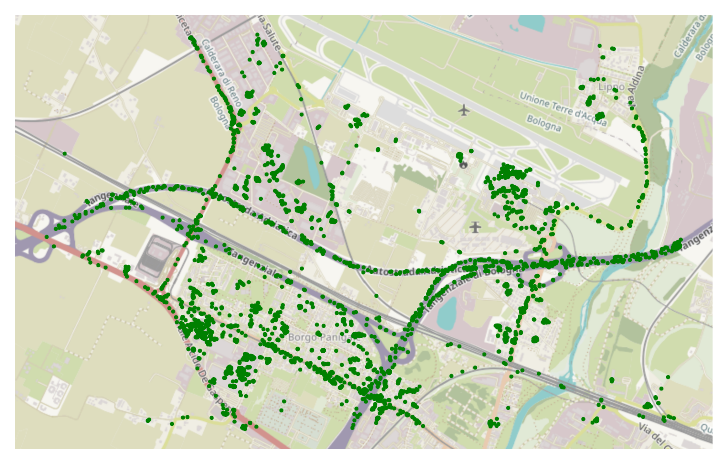}
    \vspace{2pt}
    \small (c) GMM
\end{minipage}%
\hspace{2pt}
\begin{minipage}[b]{0.19\textwidth}
    \centering
    \includegraphics[width=\linewidth,height=2.4cm]{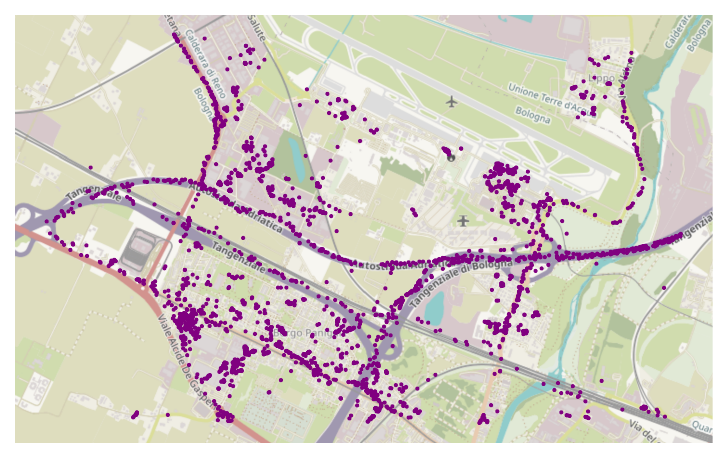}
    \vspace{2pt}
    \small (d) GAN
\end{minipage}%
\hspace{2pt}
\begin{minipage}[b]{0.19\textwidth}
    \centering
    \includegraphics[width=\linewidth,height=2.4cm]{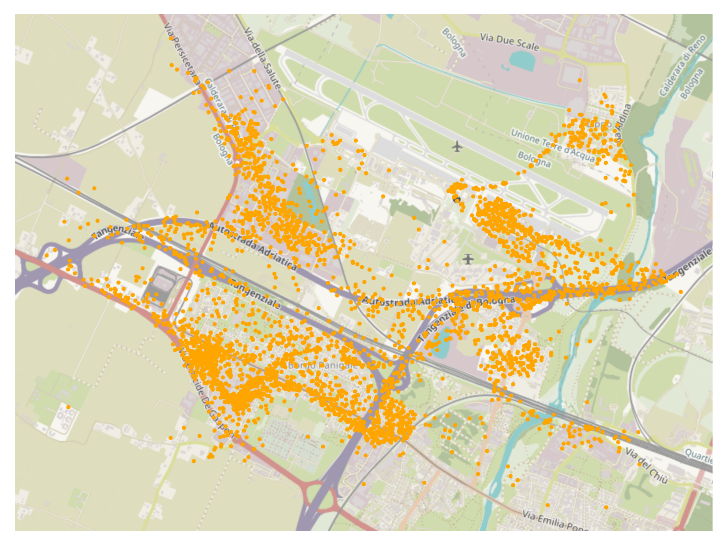}
    \vspace{2pt}
    \small (e) NF
\end{minipage}
%
\caption{Illustrative comparison of real MDT geo-located samples and synthetic locations generated by KDE, GMM, GAN, and NF for the airport scenario.
}
\label{fig:spatial_augm}
\end{figure*}
\renewcommand{\arraystretch}{1.3} 
\setlength{\tabcolsep}{2pt}       
{
\begin{table*}[t]
\color{black}
\centering
\scriptsize
\begin{tabular}{|c|c|c|c|c|c|c|c|c|c|c|c|c|} 
\hline
\multirow{2}{*}{\textbf{Model}} 
 & \multicolumn{3}{c|}{\textbf{City center}}
& \multicolumn{3}{c|}{\textbf{Stadium}} 
& \multicolumn{3}{c|}{\textbf{Airport}} 
& \multicolumn{3}{c|}{\textbf{Highway}} \\
\cline{2-13} 
& \multicolumn{1}{c|}{\textbf{Test statistic}} & \multicolumn{1}{c|}{$p$\textbf{-value}} & \multicolumn{1}{c|}{\textbf{Rej.}}
& \multicolumn{1}{c|}{\textbf{Test statistic}} & \multicolumn{1}{c|}{$p$\textbf{-value}} & \multicolumn{1}{c|}{\textbf{Rej.}}
& \multicolumn{1}{c|}{\textbf{Test statistic}} & \multicolumn{1}{c|}{$p$\textbf{-value}} & \multicolumn{1}{c|}{\textbf{Rej.}}
& \multicolumn{1}{c|}{\textbf{Test statistic}} & \multicolumn{1}{c|}{$p$\textbf{-value}} & \multicolumn{1}{c|}{\textbf{Rej.}} \\
\hline
KDE & $0.055\pm0.009$ & $0.331\pm0.227$ & 2/10 & $\textbf{0.060}\pm\textbf{0.009}$ & $0.520\pm0.224$ & 0/10 & $\textbf{0.038}\pm\textbf{0.004}$ & $0.148\pm0.112$ & 2/10 & $\textbf{0.069}\pm\textbf{0.011}$ & $0.552\pm0.196$ & 0/10 \\
GMM & $\textbf{0.051}\pm\textbf{0.009}$ & $0.441\pm0.207$ & 2/10 & $0.064\pm0.004$ & $0.632\pm0.108$ & 0/10 & $0.043\pm0.005$ & $0.066\pm0.052$ & 5/10 & $0.075\pm0.012$ & $0.449\pm0.201$ & 1/10 \\
GAN & $0.155\pm0.044$ & $0.001\pm0.000$ & 10/10 & $0.215\pm0.061$ & $0.004\pm0.006$ & 10/10 & $0.084\pm0.024$ & $0.008\pm0.014$ & 9/10 & $0.140\pm0.046$ & $0.173\pm0.251$ & 8/10 \\
NF & $0.075\pm0.012$ & $0.072\pm0.062$ & 6/10 & $0.099\pm0.020$ & $0.144\pm0.120$ & 4/10 & $0.064\pm0.012$ & $0.006\pm0.009$ & 10/10 & $0.104\pm0.024$ & $0.180\pm0.175$ & 4/10 \\
\hline
\end{tabular}
\caption{\blueR{Two-dimensional two-sample KS test statistics and \(p\)-values comparing synthetic spatial samples generated by KDE, GMM, GAN and NF against real MDT locations for each reference region. The null hypothesis (real and synthetic samples drawn from the same distribution) is rejected for \(p\text{-value}<0.05\). ``Rej.'' is the number of runs (out of ten) with $p<0.05$.}}
\label{tab:ks_test_models}
\end{table*}
}
\\The selected areas include:
\begin{itemize}
    \item The \textit{city center} area, which represents the most compact and dense region and covers the historical core of Bologna and is served by 140 distinct \acp{PCI}.
    \item The \textit{stadium} area, which corresponds to a mixed urban-rural zone covered by 104 \acp{PCI} and comprises a dense cellular deployment outside the historical core of the city.
    \item The \textit{airport} area, which is the largest zone and exhibits moderate user mobility, with service provided by 209 \acp{PCI}. While it contains the second-highest number of \acp{UE} after the city center, its users are distributed across a wider area, resulting in a reduced user density compared to the city center and stadium areas.
    \item The \textit{highway} area, which is located in the northern part of Bologna and features low user density and high user mobility, with service provided by 52 \acp{PCI}. It represents a large-scale vehicular scenario.
\end{itemize}
\blueR{
The MDT records utilized across all four reference scenarios were collected over a continuous span of one week. To evaluate the proposed augmentation framework, the dataset for each region was partitioned using a random splitting scheme into a training set (70\%), a validation set (10\%), and a held-out test set (20\%).
\\
Furthermore, a random splitting scheme over each region is adopted instead of temporal, user-based, or spatial-block splits because (i) fingerprinting and spatial density estimation require the training and test sets to share the same geographic support, and (ii) both the augmentation pipeline and the wKNN positioning estimator operate on instantaneous, memoryless MDT samples (devoid of timestamps or user identifiers) to model a static radio environment map, rather than generating or exploiting continuous time-series trajectories.
}

\section{Numerical Results}
\label{sec:results}

This section evaluates the proposed mobile data augmentation architecture both stage-by-stage and end-to-end, emphasizing its effect on fingerprinting-based positioning. We first validate each module in isolation to verify that the \emph{Spatial Features Augmentation} and the \emph{Radio Features Augmentation} behave as intended (Sec.~\ref{sec:abl_spatial} and Sec.~\ref{sec:abl_radio}, respectively). Then, we assess the impact of the full augmentation pipeline on multi-cell fingerprinting-based positioning across the four real-world MDT scenarios introduced in Sec.~\ref{sec:scenarios} (Sec.~\ref{subsec:positioning_performance}).

Our primary objective is to quantify improvements in positioning performance, while also ensuring that intermediate modules produce statistically faithful and useful synthetic data. Accordingly, module evaluations report relevant per-module metrics (i.e., multivariate two-sample \ac{KS} tests and radio prediction error metrics such as \ac{MAE}), whereas the end-to-end study focuses on positioning accuracy. All results compare against established baselines for the respective tasks, while Sec.~\ref{subsec:statistical} investigates the scaling effect of progressively augmenting the reference database (i.e., how adding more synthetic samples affects positioning).

\blueR{
For the data generation stage, the KDE bandwidth $h$ is selected via grid-search cross-validation on the validation set. The baseline generative models (GAN, CGAN, NF, ReFlow) and the MLP were trained using the Adam optimizer, with learning rates, batch sizes, and network depths tuned using the 10\% validation partition. For the end-to-end positioning evaluation, the wKNN estimator was configured with $K=3$ nearest neighbors, the Manhattan distance metric for neighbor selection, and inverse-distance weighting. These hyperparameters were kept consistent across all four evaluated scenarios. Further details are available in the provided source code.
}

\subsection{Results on Spatial Features Augmentation}
\label{sec:abl_spatial}

We compare four approaches for modeling the empirical spatial distribution of MDT samples \cite{gumble}: (i) the proposed Gaussian \ac{KDE}; (ii) a \ac{GMM} fitted via expectation-maximization with full covariances; 
(iii) a \ac{GAN} with fully-connected generator and discriminator 
trained adversarially with Adam; and (iv) a likelihood-based \ac{NF} 
trained by maximum likelihood. 
\blueR{
All methods are trained on the region-specific training partitions (comprising 70\% of the data, as detailed in Sec.~V.B) to generate synthetic 2D spatial locations. 
To ensure a balanced comparison, each generative model was sampled to produce an evaluation set whose cardinality matches that of the test set: specifically, 1291 synthetic locations in the city center, 541 in the stadium, 3540 in the airport, and 584 in the highway scenario.
}

To assess whether synthetic samples are statistically similar to real MDT locations we employ a 
\blueR{two-dimensional, two-sample \ac{KS} test, namely the Fasano--Franceschini generalization
\cite{kstest} of the \ac{KS} test.}
Using this test, we assess whether synthetic user positions generated by KDE and the competing methods are statistically consistent with the spatial distribution of positions in the original test set. 
Under the null hypothesis, the real samples and synthetic samples originate from the same distribution; we reject the null whenever the KS \(p\)-value falls below the significance level \(\alpha=0.05\). Table ~\ref{tab:ks_test_models} reports the average test statistic and the average \(p\)-value for each model and region over ten independent runs, with associated 95\% \ac{CI}\blueR{, together with the number
of runs (out of ten) in which the null hypothesis is rejected}. Fig.~\ref{fig:spatial_augm} provides an illustrative example of real and synthetic samples for one representative region.

\blueR{Overall, KDE and GMM yield the smallest test statistics in all four regions and are the models
rejected least often (Table ~\ref{tab:ks_test_models}). KDE is never rejected on stadium and highway (0/10 runs)
and is rejected in only 2/10 runs on city center and airport; its mean $p$-value stays above $\alpha$ in
every region (0.15--0.55). GMM behaves similarly on city center, stadium and highway (2/10, 0/10 and 1/10
rejections), but it is rejected in 5/10 runs on airport, where its mean $p$-value is close to the
significance threshold. Since the confidence intervals of KDE and GMM largely overlap, we do not claim that
either one is superior to the other. In contrast, the GAN is rejected in 8 to 10 out of 10 runs in all
regions, and the NF in 4/10 (stadium, highway), 6/10 (city center) and 10/10 (airport) runs, with a
statistic roughly 1.3 to 3.6 times larger than that of KDE. The relatively high mean $p$-values of the GAN
on highway should be read together with its 8/10 rejections: they reflect a large
run-to-run variability rather than a good fit.
These results indicate that KDE reproduces the joint spatial distribution of the MDT locations as well as
GMM and clearly better than GAN and NF. Small but detectable deviations remain in a few runs, mainly for
city center and airport, where the larger test sets give the test more power; we also note that failing to
reject the null hypothesis is not proof of equality between the distributions. Together with its
interpretability and light computational footprint, this supports the use of KDE as a reliable and
lightweight method for spatial sample generation in our mobile data augmentation architecture.}

\subsection{Results on Radio Features Augmentation}
\label{sec:abl_radio}

\blue{Similarly, for radio feature augmentation we compare eight methods for synthesizing per-PCI \ac{RSRP} fingerprints \cite{sun2018augmentation}: (i) the proposed spatial \blueR{1NN}-based augmentation with additive log-normal shadowing; (ii) a \ac{RF} regression trained per-PCI; (iii) a \ac{GPR} with a \ac{SE} kernel; and (iv) a \ac{GPR} with a \ac{RQ} kernel; (v) an \ac{MLP} with three hidden layers; (vi) a \ac{CGAN}; (vii) a \ac{ReFlow} model, representing state-of-the-art diffusion-based generation; and (viii) a \ac{NF} model.}
\blueR{
Furthermore, to motivate the choice of adopting a single neighbor search in our proposed KDE-1NN framework, we evaluated the impact of a varying number of nearest neighbors in $\{2, 3, 4, 5\}$ to generate radio features. 
}
All radio feature models are trained on region-specific training sets and evaluated on held-out test sets.
\blueR{
For each true spatial location $\mathbf p_i$ in the test set, we generate a synthetic radio measurement vector $\mathbf{\hat{r}}_i$ using the respective augmentation model. We then compute the \ac{MAE} between the predicted and ground-truth RSRP values. Because real-world MDT fingerprints are sparse (i.e., a user equipment only reports measurements for a subset of detectable cells), the error is exclusively computed over the valid observed measurements. Let $N_{\text{test}}$ represent the number of test fingerprints, and let $\mathcal{V}_i \subseteq \{1, \dots, N_c\}$ denote the subset of PCIs successfully detected in the true $i$-th measurement vector $\mathbf r_i$. The overall MAE is calculated as:
\begin{equation}
\text{MAE} = \frac{1}{\sum_{i=1}^{N_{\text{test}}} |\mathcal{V}_i|} \sum_{i=1}^{N_{\text{test}}} \sum_{c \in \mathcal{V}_i} \left| \mathbf r_{i,c} - \mathbf{\hat{r}}_{i,c} \right|
\end{equation}
where $\mathbf r_{i,c}$ and $\mathbf{\hat{r}}_{i,c}$ are the true and synthetically generated RSRP values for cell $c$ at the $i$-th test location, respectively.}  This evaluation is conducted independently for each geographic region. The reported values in Table ~\ref{tab:mae_models} represent the average region-level \ac{MAE} over ten independent runs, with associated 95\% \ac{CI}.

The results demonstrate that all models achieve relatively low \ac{MAE} values, indicating effective RSRP estimation across diverse urban and peri-urban regions. The \blueR{1NN} model consistently yields the lowest \ac{MAE} across all regions, particularly excelling in the airport and city center zones, which contain the highest number of \acp{UE}. In contrast, the \ac{GPR} models show slightly worse overall performance. GPR (\ac{RQ}) consistently outperforms the \ac{GPR} (SE) across all regions, as it incorporates a more flexible kernel capable of modeling multi-scale signal variations. \ac{RF} exhibits moderate performance, with accuracy generally falling between \blueR{1NN} and \ac{GPR}. 
\blueR{While \ac{GPR} explicitly models global spatial correlation, it tends to over-smooth the high-frequency spatial variations caused by local shadowing in sparse data regimes. Conversely, the 1NN approach combined with shadow-fading noise inherently preserves these irregular and local propagation characteristics better.}

\blue{Moreover, the experimental results reveal a relevant trend: lightweight, instance-based methods (specifically \blueR{1NN}) consistently outperform complex deep generative architectures (such as \ac{CGAN}, \ac{ReFlow}, and \ac{NF}) in \ac{RSRP} estimation. This performance gap is primarily driven by the context-agnostic nature of the problem, where the input space is restricted to low-dimensional spatial coordinates (latitude, longitude) without auxiliary environmental metadata. While deep learning models typically excel at extracting hierarchical features from high-dimensional structured data, they often struggle to model the high-frequency spatial variations inherent to radio propagation (e.g., local shadowing and multipath) when constrained to such sparse inputs, which is a phenomenon often linked to the spectral bias of neural networks~\cite{rahaman2019spectral}. In contrast, \blueR{1NN} operates as a non-parametric learner with a strong local inductive bias. By avoiding assumptions about global function smoothness, it effectively captures the localized and irregular nature of the radio environment, proving that in the absence of geometric features, local interpolation remains superior to deep functional approximation.}
\blueR{
Furthermore, Table \ref{tab:mae_models} shows that, among the evaluated neighbor-count variants, 1NN consistently yields the lowest MAE across all scenarios. Increasing the number of neighbors leads to a progressive performance degradation; this occurs because the spatial averaging of multiple neighbors in complex outdoor environments tends to over-smooth the RSRP values.
} 

\begin{table}[t]
\centering
\vspace{0.2cm}
\renewcommand{\arraystretch}{1.4}
\resizebox{\columnwidth}{!}{
\begin{tabular}{|>{\centering\arraybackslash}m{1.9cm}|c|c|c|c|}
\hline
\textbf{Model} & \textbf{City center} & \textbf{Stadium} & \textbf{Airport} & \textbf{Highway} \\
\hline
\blueR{1NN} & \textbf{0.52$\pm$0.01} & \textbf{0.85$\pm$0.01} & \textbf{0.48$\pm$0.01} & \textbf{2.11$\pm$0.01} \\

\blueR{2NN} & \blueR{0.55$\pm$0.01} & \blueR{0.87$\pm$0.02} & \blueR{0.48$\pm$0.01} & \blueR{2.22$\pm$0.09} \\

\blueR{3NN} & \blueR{0.56$\pm$0.01} & \blueR{0.90$\pm$0.02} & \blueR{0.48$\pm$0.01} & \blueR{2.22$\pm$0.10} \\

\blueR{4NN} & \blueR{0.57$\pm$0.01} & \blueR{0.93$\pm$0.02} & \blueR{0.49$\pm$0.01} & \blueR{2.24$\pm$0.08} \\

\blueR{5NN} & \blueR{0.60$\pm$0.02} & \blueR{0.99$\pm$0.02} & \blueR{0.50$\pm$0.01} & \blueR{2.27$\pm$0.07} \\

\ac{RF}       & 0.59$\pm$0.01 & 0.98$\pm$0.01 & 0.52$\pm$0.02 & 2.42$\pm$0.01 \\

\ac{GPR} (\ac{SE})       & 0.71$\pm$0.00 & 1.05$\pm$0.01 & 0.66$\pm$0.01 & 2.86$\pm$0.03 \\

\ac{GPR} (\ac{RQ})       & 0.61$\pm$0.08 & 0.95$\pm$0.02 & 0.50$\pm$0.18 & 2.43$\pm$0.01 \\

\blue{\ac{MLP}}   & \blue{0.80$\pm$0.01} & \blue{1.17$\pm$0.02} & \blue{0.59$\pm$0.01} & \blue{2.34$\pm$0.04}\\

\blue{\ac{CGAN}}   & \blue{0.81$\pm$0.01} & \blue{1.17$\pm$0.02} & \blue{0.60$\pm$0.01} & \blue{2.31$\pm$0.03}\\

\blue{\ac{ReFlow}}  & \blue{1.11$\pm$0.03} & \blue{1.44$\pm$0.02} & \blue{0.92$\pm$0.01} & \blue{2.62$\pm$0.03} \\

\blue{\ac{NF}}   & \blue{1.74$\pm$0.22} & \blue{2.02$\pm$0.16} & \blue{1.22$\pm$0.53} & \blue{4.20$\pm$0.72} \\
\hline
\end{tabular}
}
\caption{\blue{\ac{MAE} of \ac{RSRP} prediction for considered approaches in each reference region.}}

\label{tab:mae_models}
\end{table}

\subsection{Results on Multi-cell Fingerprinting-based Positioning}
\label{subsec:positioning_performance}

In this section, we evaluate the end-to-end impact of the proposed mobile data augmentation architecture on multi-cell fingerprinting-based positioning. Starting from the region-specific reference database, we compare positioning performance obtained with the original operator-collected fingerprints against databases augmented with synthetic samples produced by the spatial and radio augmentation stages. The positioning algorithm under test is the \ac{wKNN} estimator introduced in Sec.~\ref{sec:positioning}; results are reported for all four MDT reference scenarios. 

We compare our proposed two-stage architecture \ac{KDE}-\blueR{\ac{1NN}} against these alternative two-stage configurations: \ac{KDE}-\ac{RF}, \ac{KDE}-\ac{GPR} (\ac{SE}/\ac{RQ}), \ac{KDE}-\ac{MLP}, \ac{KDE}-\ac{CGAN}, \ac{KDE}-\ac{ReFlow}, and \ac{KDE}-\ac{NF}.\blueR{
Furthermore, we evaluate the end-to-end positioning performance when varying the number of nearest neighbors in $\{2, 3, 4, 5\}$. 
}For each MDT scenario, we evaluate a set of augmentation rates \(A\in\{1,5,10,20,30\}\), where \(A\) denotes the multiplicative increase of the fingerprinting database relative to the original (\(A=1\) corresponds to no augmentation). For every combination of scenario, augmentation rate and architecture, we perform ten independent runs. In the remainder of this section, we report average \ac{MedAE} values for positioning with 95\% \ac{CI}. \green{We quantify the positioning gain at augmentation rate $A$ as the relative improvement with respect to the non-augmented baseline, that is, $\Delta(A) = \bigl(\mathrm{MedAE}_{A=1} - \mathrm{MedAE}_{A}\bigr)/\mathrm{MedAE}_{A=1} \times 100\%$.}

\begin{enumerate}
\item \textit{City Center Area} (Table~\ref{tab:aug_comparison_citycenter}).

In the city center scenario, which features the highest measurement density and the smallest spatial extent, the \ac{wKNN} estimator applied to the original MDT baseline attains a median positioning error of $20.08\,$m. With the proposed \ac{KDE}-\blueR{\ac{1NN}} pipeline and an augmentation factor $A=20$, the median error decreases to $17.80\,$m, corresponding to a relative improvement of $\green{\Delta(20) \approx 11.4\%}$. Although the absolute reduction ($\approx2.3\,$m) is modest, it is noteworthy given the strong baseline performance. Moreover, \ac{KDE}-\blueR{\ac{1NN}} consistently outperforms the alternative two-stage configurations, which yield smaller or no gains; in particular, the \ac{KDE}-\ac{RF} variant may be detrimental, since RF-based RSRP predictions can produce augmented fingerprints that degrade positioning accuracy.
    
\item \textit{Stadium Area} (Table~\ref{tab:aug_comparison_stadium}).

\begin{table*}[t]
\centering
\renewcommand{\arraystretch}{1.25}
\setlength{\tabcolsep}{15pt}
\begin{tabular}{lccccc}
\hline
\textbf{    Model} & $\times 1$ \textit{(Not Augmented)} & $\times 5$ & $\times 10$ & $\times 20$ & $\times 30$ \\
\hline
\ac{KDE}-\blueR{\ac{1NN}} & 20.08$\pm$0.64 & 19.59$\pm$1.81 & 18.19$\pm$0.56 & \textbf{17.80$\pm$0.59} & 18.31$\pm$0.36 \\
{\color{black}\ac{KDE}-2NN} & {\color{black}20.08$\pm$0.64} & {\color{black}19.44$\pm$1.06} & {\color{black}18.88$\pm$0.22} & {\color{black}18.83$\pm$0.32} & {\color{black}19.06$\pm$0.44} \\
{\color{black}\ac{KDE}-3NN} & {\color{black}20.08$\pm$0.64} & {\color{black}21.59$\pm$0.26} & {\color{black}21.66$\pm$0.46} & {\color{black}21.42$\pm$0.33} & {\color{black}21.47$\pm$0.37} \\
{\color{black}\ac{KDE}-4NN} & {\color{black}20.08$\pm$0.64} & {\color{black}22.58$\pm$0.31} & {\color{black}22.32$\pm$0.30} & {\color{black}21.94$\pm$0.36} & {\color{black}22.07$\pm$0.23} \\
{\color{black}\ac{KDE}-5NN} & {\color{black}20.08$\pm$0.64} & {\color{black}22.02$\pm$0.37} & {\color{black}21.63$\pm$0.29} & {\color{black}21.57$\pm$0.41} & {\color{black}21.61$\pm$0.32} \\
\ac{KDE}-\ac{RF} & 20.08$\pm$0.64 & 24.03$\pm$0.20 & 23.32$\pm$0.24 & 20.15$\pm$0.20 & 22.33$\pm$0.21 \\
\ac{KDE}-\ac{GPR} & 20.08$\pm$0.64 & 21.29$\pm$0.15 & 20.68$\pm$0.16 & 19.38$\pm$0.19 & 19.28$\pm$0.23 \\
\ac{KDE}-\ac{GPR}(\ac{RQ}) & 20.08$\pm$0.64 & 20.61$\pm$0.18 & 20.14$\pm$0.17 & 18.71$\pm$0.14 & 19.07$\pm$0.16 \\
\blue{\ac{KDE}-\ac{MLP}} & \blue{20.08$\pm$0.64} & \blue{25.80$\pm$0.14} & \blue{24.10$\pm$0.14} & \blue{22.80$\pm$0.19} & \blue{22.00$\pm$0.13} \\
\blue{\ac{KDE}-\ac{CGAN}} & \blue{20.08$\pm$0.64} & \blue{30.50$\pm$0.57} & \blue{28.40$\pm$0.54} & \blue{26.50$\pm$0.45} & \blue{27.80$\pm$0.41} \\
\blue{\ac{KDE}-\ac{ReFlow}} & \blue{20.08$\pm$0.64} & \blue{24.50$\pm$0.16} & \blue{23.80$\pm$0.28} & \blue{20.85$\pm$0.56} & \blue{21.50$\pm$1.17} \\
\blue{\ac{KDE}-\ac{NF}} & \blue{20.08$\pm$0.64} & \blue{40.88$\pm$1.12} & \blue{39.10$\pm$0.37} & \blue{37.50$\pm$0.32} & \blue{37.80$\pm$1.24} \\
\hline
\end{tabular}
\vspace{2pt}
\caption{\blue{Impact of data augmentation rates on positioning across increasing synthetic data ratios in the \emph{city center} scenario.}}
\label{tab:aug_comparison_citycenter}
\end{table*}

\begin{table*}[t]
\centering
\renewcommand{\arraystretch}{1.25}
\setlength{\tabcolsep}{15pt}
\begin{tabular}{lccccc}
\hline
\textbf{Model} & $\times 1$ \textit{(Not Augmented)} & $\times 5$ & $\times 10$ & $\times 20$ & $\times 30$ \\
\hline
\ac{KDE}-\blueR{\ac{1NN}} & 36.70$\pm$0.57 & 31.70$\pm$0.71 & 30.72$\pm$0.69 & \textbf{28.32$\pm$0.81} & 28.85$\pm$0.38 \\
{\color{black}\ac{KDE}-2NN} & {\color{black}36.70$\pm$0.57} & {\color{black}33.87$\pm$1.57} & {\color{black}33.24$\pm$2.42} & {\color{black}32.13$\pm$1.54} & {\color{black}32.30$\pm$1.67} \\
{\color{black}\ac{KDE}-3NN} & {\color{black}36.70$\pm$0.57} & {\color{black}35.18$\pm$1.15} & {\color{black}33.32$\pm$1.14} & {\color{black}34.00$\pm$0.99} & {\color{black}34.56$\pm$1.20} \\
{\color{black}\ac{KDE}-4NN} & {\color{black}36.70$\pm$0.57} & {\color{black}40.12$\pm$1.18} & {\color{black}38.65$\pm$1.81} & {\color{black}37.41$\pm$1.61} & {\color{black}37.06$\pm$1.79} \\
{\color{black}\ac{KDE}-5NN} & {\color{black}36.70$\pm$0.57} & {\color{black}43.29$\pm$1.83} & {\color{black}42.11$\pm$0.73} & {\color{black}41.59$\pm$1.52} & {\color{black}42.03$\pm$0.71} \\
\ac{KDE}-\ac{RF} & 36.70$\pm$0.57 & 40.15$\pm$0.61 & 39.24$\pm$0.52 & 36.86$\pm$0.38 & 36.32$\pm$0.29 \\
\ac{KDE}-\ac{GPR} & 36.70$\pm$0.57 & 39.05$\pm$0.30 & 38.40$\pm$0.24 & 34.73$\pm$0.29 & 32.90$\pm$0.13 \\
\ac{KDE}-\ac{GPR}(\ac{RQ}) & 36.70$\pm$0.57 & 37.99$\pm$0.14 & 33.48$\pm$0.18 & 30.61$\pm$0.28 & 30.05$\pm$0.11 \\
\blue{\ac{KDE}-\ac{MLP}} & \blue{36.70$\pm$0.57} & \blue{39.50$\pm$0.23} & \blue{38.79$\pm$0.22} & \blue{35.80$\pm$0.20} & \blue{37.10$\pm$0.25} \\
\blue{\ac{KDE}-\ac{CGAN}} & \blue{36.70$\pm$0.57} & \blue{48.50$\pm$0.58} & \blue{46.80$\pm$0.49} & \blue{45.20$\pm$0.59} & \blue{46.00$\pm$0.77} \\
\blue{\ac{KDE}-\ac{ReFlow}} & \blue{36.70$\pm$0.57} & \blue{41.20$\pm$0.39} & \blue{40.50$\pm$0.28} & \blue{38.50$\pm$0.32} & \blue{38.90$\pm$1.24} \\
\blue{\ac{KDE}-\ac{NF}} & \blue{36.70$\pm$0.57} & \blue{55.34$\pm$0.57} & \blue{55.60$\pm$0.47} & \blue{53.51$\pm$0.81} & \blue{54.00$\pm$0.82} \\
\hline
\end{tabular}
\vspace{2pt}
\caption{\blue{Impact of data augmentation rates on positioning across increasing synthetic data ratios in the \emph{stadium} scenario.}}
\label{tab:aug_comparison_stadium}
\end{table*}

\begin{table*}[t]
\centering
\renewcommand{\arraystretch}{1.25}
\setlength{\tabcolsep}{15pt}
\begin{tabular}{lccccc}
\hline
\textbf{Model} & $\times 1$ \textit{(Not Augmented)} & $\times 5$ & $\times 10$ & $\times 20$ & $\times 30$ \\
\hline
\ac{KDE}-\blueR{\ac{1NN}} & 72.04$\pm$0.64 & 70.34$\pm$1.20 & 69.27$\pm$1.16 & \textbf{68.54$\pm$0.87} & 69.39$\pm$0.65 \\
{\color{black}\ac{KDE}-2NN} & {\color{black}72.04$\pm$0.64} & {\color{black}72.14$\pm$0.17} & {\color{black}71.90$\pm$0.67} & {\color{black}71.74$\pm$0.27} & {\color{black}71.71$\pm$0.16} \\
{\color{black}\ac{KDE}-3NN} & {\color{black}72.04$\pm$0.64} & {\color{black}74.00$\pm$0.32} & {\color{black}73.90$\pm$0.25} & {\color{black}73.65$\pm$0.33} & {\color{black}73.68$\pm$0.22} \\
{\color{black}\ac{KDE}-4NN} & {\color{black}72.04$\pm$0.64} & {\color{black}74.46$\pm$0.28} & {\color{black}74.27$\pm$0.42} & {\color{black}74.02$\pm$0.79} & {\color{black}73.94$\pm$0.28} \\
{\color{black}\ac{KDE}-5NN} & {\color{black}72.04$\pm$0.64} & {\color{black}77.06$\pm$0.67} & {\color{black}76.79$\pm$0.43} & {\color{black}75.76$\pm$0.90} & {\color{black}74.76$\pm$0.81} \\
\ac{KDE}-\ac{RF} & 72.04$\pm$0.64 & 86.26$\pm$1.26 & 87.87$\pm$1.14 & 88.91$\pm$1.14 & 91.78$\pm$1.52 \\
\ac{KDE}-\ac{GPR} & 72.04$\pm$0.64 & 83.43$\pm$2.24 & 81.26$\pm$2.85 & 88.29$\pm$1.23 & 88.79$\pm$1.64 \\
\ac{KDE}-\ac{GPR}(\ac{RQ}) & 72.04$\pm$0.64 & 79.21$\pm$1.12 & 86.04$\pm$1.82 & 80.45$\pm$2.04 & 85.71$\pm$1.40 \\
\blue{\ac{KDE}-\ac{MLP}} & \blue{72.04$\pm$0.64} & \blue{87.50$\pm$0.72} & \blue{88.10$\pm$0.62} & \blue{85.50$\pm$1.39} & \blue{88.20$\pm$1.59} \\
\blue{\ac{KDE}-\ac{CGAN}} & \blue{72.04$\pm$0.64} & \blue{98.50$\pm$0.70} & \blue{104.20$\pm$0.87} & \blue{103.10$\pm$2.19} & \blue{101.40$\pm$1.60} \\
\blue{\ac{KDE}-\ac{ReFlow}} & \blue{72.04$\pm$0.64} & \blue{94.20$\pm$0.93} & \blue{93.80$\pm$2.41} & \blue{91.20$\pm$1.52} & \blue{92.50$\pm$1.30} \\
\blue{\ac{KDE}-\ac{NF}} & \blue{72.04$\pm$0.64} & \blue{110.34$\pm$1.66} & \blue{111.27$\pm$2.62} & \blue{109.54$\pm$1.49} & \blue{110.39$\pm$0.39} \\
\hline
\end{tabular}
\vspace{2pt}
\caption{\blue{Impact of data augmentation rates on positioning across increasing synthetic data ratios in the \emph{airport} scenario.}}
\label{tab:aug_comparison_airport}
\end{table*}
\begin{table*}[!t]
\centering
\renewcommand{\arraystretch}{1.25}
\setlength{\tabcolsep}{15pt}
\begin{tabular}{lccccc}
\hline
\textbf{   Model} & $\times 1$ \textit{(Not Augmented)} & $\times 5$ & $\times 10$ & $\times 20$ & $\times 30$ \\
\hline
\ac{KDE}-\blueR{\ac{1NN}} & 115.51$\pm$0.86 & 95.23$\pm$0.34 & 82.41$\pm$0.19 & 82.97$\pm$0.22 & \textbf{79.82$\pm$0.31} \\
{\color{black}\ac{KDE}-2NN} & {\color{black}115.51$\pm$0.86} & {\color{black}119.55$\pm$1.87} & {\color{black}118.66$\pm$2.01} & {\color{black}114.45$\pm$1.95} & {\color{black}111.50$\pm$1.30} \\
{\color{black}\ac{KDE}-3NN} & {\color{black}115.51$\pm$0.86} & {\color{black}121.27$\pm$1.81} & {\color{black}121.32$\pm$1.46} & {\color{black}119.50$\pm$1.89} & {\color{black}118.28$\pm$1.93} \\
{\color{black}\ac{KDE}-4NN} & {\color{black}115.51$\pm$0.86} & {\color{black}125.61$\pm$1.77} & {\color{black}123.89$\pm$1.48} & {\color{black}121.39$\pm$1.98} & {\color{black}120.34$\pm$1.48} \\
{\color{black}\ac{KDE}-5NN} & {\color{black}115.51$\pm$0.86} & {\color{black}132.22$\pm$1.57} & {\color{black}131.77$\pm$1.58} & {\color{black}130.91$\pm$1.49} & {\color{black}128.31$\pm$2.00} \\
\ac{KDE}-\ac{RF} & 115.51$\pm$0.86 & 124.04$\pm$2.47 & 120.34$\pm$1.69 & 116.27$\pm$1.51 & 116.23$\pm$1.26 \\
\ac{KDE}-\ac{GPR} & 115.51$\pm$0.86 & 117.65$\pm$1.46 & 109.35$\pm$1.13 & 104.35$\pm$1.21 & 103.21$\pm$1.05 \\
\ac{KDE}-\ac{GPR}(\ac{RQ}) & 115.51$\pm$0.86 & 109.25$\pm$1.41 & 99.32$\pm$0.97 & 94.35$\pm$1.43 & 91.74$\pm$1.24 \\
\blue{\ac{KDE}-\ac{MLP}} & \blue{115.51$\pm$0.86} & \blue{125.80$\pm$0.93} & \blue{118.50$\pm$1.20} & \blue{114.50$\pm$1.18} & \blue{113.80$\pm$1.04} \\
\blue{\ac{KDE}-\ac{CGAN}} & \blue{115.51$\pm$0.86} & \blue{145.20$\pm$1.91} & \blue{141.10$\pm$1.36} & \blue{135.28$\pm$1.32} & \blue{135.50$\pm$1.64} \\
\blue{\ac{KDE}-\ac{ReFlow}} & \blue{115.51$\pm$0.86} & \blue{130.50$\pm$1.14} & \blue{125.80$\pm$1.56} & \blue{119.40$\pm$1.23} & \blue{121.10$\pm$1.33} \\
\blue{\ac{KDE}-\ac{NF}} & \blue{115.51$\pm$0.86} & \blue{168.03$\pm$1.14} & \blue{171.01$\pm$1.09} & \blue{171.17$\pm$0.92} & \blue{172.02$\pm$1.30} \\
\hline
\end{tabular}
\vspace{2pt}
\caption{\blue{Impact of data augmentation rates on positioning across increasing synthetic data ratios in the \emph{highway} scenario.}}
\label{tab:aug_comparison_highway}
\end{table*}
In the stadium scenario, characterized by a high measurement density but a very confined spatial extent and an intermediate number of \acp{PCI}, the proposed \ac{KDE}-\blueR{\ac{1NN}} pipeline yields substantial gains. With augmentation factor \(A=20\), the median positioning error drops from \(36.70\)\,m to \(28.32\)\,m, an absolute improvement of \(8.38\)\,m and a relative reduction of $\green{\Delta(20) \approx 23\%}$. By contrast, the alternative two-stage configurations provide only marginal or no improvements across augmentation levels.

\item \textit{Airport Area} (Table~\ref{tab:aug_comparison_airport}).

In the airport scenario, which corresponds to the largest area in our study and the one served by the highest number of \acp{PCI}, the propagation environment and spatial variability are markedly more complex. Under these conditions, the \ac{KDE}-\blueR{\ac{1NN}} pipeline yields only a modest improvement of $\green{\Delta(20) \approx 4\%}$, while other alternative augmentation schemes actually degrade positioning performance. These findings underline that simply increasing the quantity of synthetic samples is not sufficient: augmentation must preserve the statistical and spatial structure of the radio field, otherwise imperfect radio predictors can introduce distortions that harm positioning accuracy.

\item \textit{Highway Area} (Table~\ref{tab:aug_comparison_highway}).

In the highway scenario, where UE measurements are highly concentrated along the linear roadway (i.e., samples cluster on the highway) but sparse elsewhere across the largest geographic extent and subject to high mobility, augmentation delivers the largest benefit. With the proposed \ac{KDE}-\blueR{\ac{1NN}} pipeline and augmentation factor \(A=30\), the median positioning error falls from \(115.51\)\,m to \(79.82\)\,m, a relative reduction of $\green{\Delta(30) \approx 30\%}$. This substantial improvement indicates that realistic augmentation can compensate for the anisotropic and sparse spatial sampling typical of highway measurements; by contrast, the alternative two-stage configurations produce markedly smaller gains and are consistently outperformed by \ac{KDE}-\blueR{\ac{1NN}} across augmentation levels.

\end{enumerate}
These findings lead to \blue{five} relevant conclusions: 
(i) the \ac{KDE}-\blueR{\ac{1NN}} pipeline consistently improves positioning accuracy across all evaluated scenarios, with the largest relative gains in spatially sparse or structurally complex regions; 
(ii) the alternative two-stage baselines fail to match the proposed approach across the tested augmentation rates;
(iii) \blueR{consistent with the RSRP prediction results, the 1NN configuration provides the most substantial reduction in positioning error at every evaluated augmentation rate across nearest neighbors approaches;}
(iv) performance gains saturate beyond a region-specific augmentation factor, so adding more synthetic samples does not by itself guarantee better accuracy, and poorly matched synthetic fingerprints can even degrade it; \blue{and 
(v) in high-density scenarios (e.g., \emph{city center} or \emph{stadium}), the saturation in terms of positioning performance occurs at low augmentation rates and converges to low values, as UE density is already high in the training set, as shown in Table~\ref{tab:original_density}. Conversely, in the \emph{highway} scenario, characterized by high mobility and extremely low spatial density, the impact of data sparsity is significant. Consequently, this region requires much higher augmentation rates before saturation is reached and overall positioning algorithms perform worse.}

\subsection{Statistical Model Comparison}
\label{subsec:statistical}

\noindent
\begin{figure*}[t]
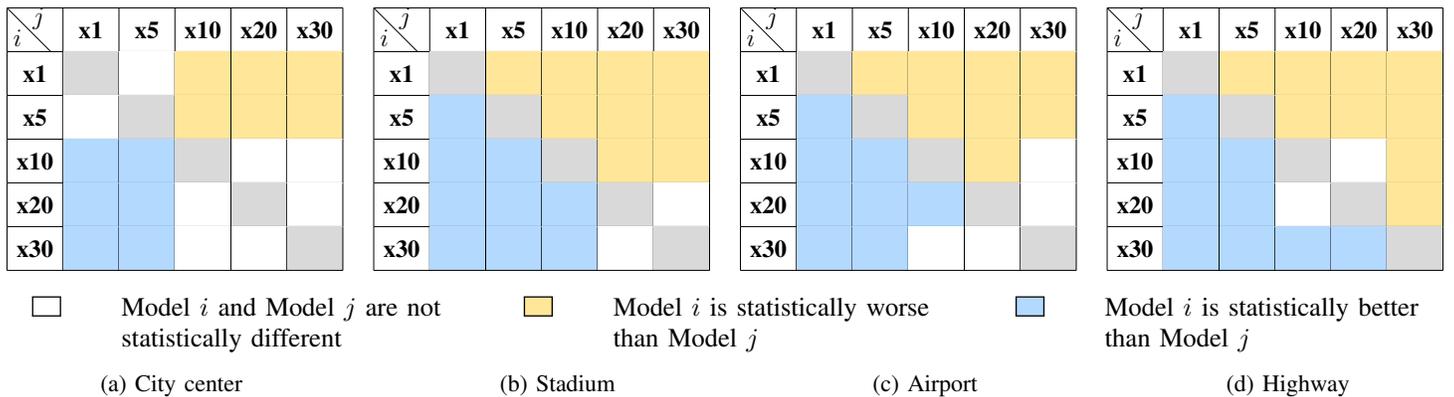

\centering
\resizebox{\textwidth}{!}{
\begin{tabular}{@{}c@{\hskip 30pt}c@{\hskip 30pt}c@{\hskip 30pt}c@{}}
\begin{tabular}{|p{1.5cm}|p{1.5cm}|p{1.5cm}|p{1.5cm}|p{1.5cm}|p{1.5cm}|p{1.5cm}|}
\hline
\diagbox[width=1.5cm, height=1.5cm]{\Huge $i$}{\Huge $j$} 
& \parbox[c][1.5cm][c]{1.5cm}{\centering \Huge \textbf{x1}} 
& \parbox[c][1.5cm][c]{1.5cm}{\centering \Huge \textbf{x5}} 
& \parbox[c][1.5cm][c]{1.5cm}{\centering \Huge \textbf{x10}} 
& \parbox[c][1.5cm][c]{1.5cm}{\centering \Huge \textbf{x20}} 
& \parbox[c][1.5cm][c]{1.5cm}{\centering \Huge \textbf{x30}}  \\
\hline
\parbox[c][1.5cm][c]{1.5cm}{\centering \Huge \textbf{x1}}  & \cellcolor{diaggray}\parbox[c][1.5cm][c]{1.5cm}{\centering \Huge } & \cellcolor{neutral}\parbox[c][1.5cm][c]{1.5cm}{\centering \Huge } & \cellcolor{worse}\parbox[c][1.5cm][c]{1.5cm}{\centering \Huge } & \cellcolor{worse}\parbox[c][1.5cm][c]{1.5cm}{\centering \Huge } & \cellcolor{worse}\parbox[c][1.5cm][c]{1.5cm}{\centering \Huge } \\
\hline
\parbox[c][1.5cm][c]{1.5cm}{\centering \Huge \textbf{x5}}  & \cellcolor{neutral}\parbox[c][1.5cm][c]{1.5cm}{\centering \Huge } & \cellcolor{diaggray}\parbox[c][1.5cm][c]{1.5cm}{\centering \Huge } & \cellcolor{worse}\parbox[c][1.5cm][c]{1.5cm}{\centering \Huge } & \cellcolor{worse}\parbox[c][1.5cm][c]{1.5cm}{\centering \Huge } & \cellcolor{worse}\parbox[c][1.5cm][c]{1.5cm}{\centering \Huge }\\
\hline
\parbox[c][1.5cm][c]{1.5cm}{\centering \Huge \textbf{x10}} & \cellcolor{better}\parbox[c][1.5cm][c]{1.5cm}{\centering \Huge } & \cellcolor{better}\parbox[c][1.5cm][c]{1.5cm}{\centering \Huge } & \cellcolor{diaggray}\parbox[c][1.5cm][c]{1.5cm}{\centering \Huge } & \cellcolor{neutral}\parbox[c][1.5cm][c]{1.5cm}{\centering \Huge } & \cellcolor{neutral}\parbox[c][1.5cm][c]{1.5cm}{\centering \Huge } \\
\hline
\parbox[c][1.5cm][c]{1.5cm}{\centering \Huge \textbf{x20}} & \cellcolor{better}\parbox[c][1.5cm][c]{1.5cm}{\centering \Huge } & \cellcolor{better}\parbox[c][1.5cm][c]{1.5cm}{\centering \Huge } & \cellcolor{neutral}\parbox[c][1.5cm][c]{1.5cm}{\centering \Huge ~} & \cellcolor{diaggray}\parbox[c][1.5cm][c]{1.5cm}{\centering \Huge } & \cellcolor{neutral}\parbox[c][1.5cm][c]{1.5cm}{\centering \Huge } \\
\hline
\parbox[c][1.5cm][c]{1.5cm}{\centering \Huge \textbf{x30}} & \cellcolor{better}\parbox[c][1.5cm][c]{1.5cm}{\centering \Huge } & \cellcolor{better}\parbox[c][1.5cm][c]{1.5cm}{\centering \Huge } & \cellcolor{neutral}\parbox[c][1.5cm][c]{1.5cm}{\centering \Huge } & \cellcolor{neutral}\parbox[c][1.5cm][c]{1.5cm}{\centering \Huge } & \cellcolor{diaggray}\parbox[c][1.5cm][c]{1.5cm}{\centering \Huge }\\
\hline
\end{tabular}
& 
\begin{tabular}{|p{1.5cm}|p{1.5cm}|p{1.5cm}|p{1.5cm}|p{1.5cm}|p{1.5cm}|}
\hline
\diagbox[width=1.5cm, height=1.5cm]{\Huge $i$}{\Huge $j$} 
& \parbox[c][1.5cm][c]{1.5cm}{\centering \Huge \textbf{x1}} 
& \parbox[c][1.5cm][c]{1.5cm}{\centering \Huge \textbf{x5}} 
& \parbox[c][1.5cm][c]{1.5cm}{\centering \Huge \textbf{x10}} 
& \parbox[c][1.5cm][c]{1.5cm}{\centering \Huge \textbf{x20}} 
& \parbox[c][1.5cm][c]{1.5cm}{\centering \Huge \textbf{x30}} \\ 
\hline
\parbox[c][1.5cm][c]{1.5cm}{\centering \Huge \textbf{x1}}  & \cellcolor{diaggray}\parbox[c][1.5cm][c]{1.5cm}{\centering \Huge } & \cellcolor{worse}\parbox[c][1.5cm][c]{1.5cm}{\centering \Huge} & \cellcolor{worse}\parbox[c][1.5cm][c]{1.5cm}{\centering \Huge} & \cellcolor{worse}\parbox[c][1.5cm][c]{1.5cm}{\centering \Huge} & \cellcolor{worse}\parbox[c][1.5cm][c]{1.5cm}{\centering \Huge} \\ 
\hline
\parbox[c][1.5cm][c]{1.5cm}{\centering \Huge \textbf{x5}}  & \cellcolor{better}\parbox[c][1.5cm][c]{1.5cm}{\centering \Huge } & \cellcolor{diaggray}\parbox[c][1.5cm][c]{1.5cm}{\centering \Huge } & \cellcolor{worse}\parbox[c][1.5cm][c]{1.5cm}{\centering \Huge } & \cellcolor{worse}\parbox[c][1.5cm][c]{1.5cm}{\centering \Huge } & \cellcolor{worse}\parbox[c][1.5cm][c]{1.5cm}{\centering \Huge } \\ 
\hline
\parbox[c][1.5cm][c]{1.5cm}{\centering \Huge \textbf{x10}} & \cellcolor{better}\parbox[c][1.5cm][c]{1.5cm}{\centering \Huge } & \cellcolor{better}\parbox[c][1.5cm][c]{1.5cm}{\centering \Huge } & \cellcolor{diaggray}\parbox[c][1.5cm][c]{1.5cm}{\centering \Huge } & \cellcolor{worse}\parbox[c][1.5cm][c]{1.5cm}{\centering \Huge } & \cellcolor{worse}\parbox[c][1.5cm][c]{1.5cm}{\centering \Huge } \\ 
\hline
\parbox[c][1.5cm][c]{1.5cm}{\centering \Huge \textbf{x20}} & \cellcolor{better}\parbox[c][1.5cm][c]{1.5cm}{\centering \Huge } & \cellcolor{better}\parbox[c][1.5cm][c]{1.5cm}{\centering \Huge } & \cellcolor{better}\parbox[c][1.5cm][c]{1.5cm}{\centering \Huge } & \cellcolor{diaggray}\parbox[c][1.5cm][c]{1.5cm}{\centering \Huge} & \cellcolor{neutral}\parbox[c][1.5cm][c]{1.5cm}{\centering \Huge ~} \\ 
\hline
\parbox[c][1.5cm][c]{1.5cm}{\centering \Huge \textbf{x30}} & \cellcolor{better}\parbox[c][1.5cm][c]{1.5cm}{\centering \Huge } & \cellcolor{better}\parbox[c][1.5cm][c]{1.5cm}{\centering \Huge } & \cellcolor{better}\parbox[c][1.5cm][c]{1.5cm}{\centering \Huge } & \cellcolor{neutral}\parbox[c][1.5cm][c]{1.5cm}{\centering \Huge ~} & \cellcolor{diaggray}\parbox[c][1.5cm][c]{1.5cm}{\centering \Huge } \\ 
\hline
\end{tabular}
& 
\begin{tabular}{|p{1.5cm}|p{1.5cm}|p{1.5cm}|p{1.5cm}|p{1.5cm}|p{1.5cm}|p{1.5cm}|}
\hline
\diagbox[width=1.5cm, height=1.5cm]{\Huge $i$}{\Huge $j$} 
& \parbox[c][1.5cm][c]{1.5cm}{\centering \Huge \textbf{x1}} 
& \parbox[c][1.5cm][c]{1.5cm}{\centering \Huge \textbf{x5}} 
& \parbox[c][1.5cm][c]{1.5cm}{\centering \Huge \textbf{x10}} 
& \parbox[c][1.5cm][c]{1.5cm}{\centering \Huge \textbf{x20}} 
& \parbox[c][1.5cm][c]{1.5cm}{\centering \Huge \textbf{x30}}  \\
\hline
\parbox[c][1.5cm][c]{1.5cm}{\centering \Huge \textbf{x1}}  & \cellcolor{diaggray}\parbox[c][1.5cm][c]{1.5cm}{\centering \Huge } & \cellcolor{worse}\parbox[c][1.5cm][c]{1.5cm}{\centering \Huge } & \cellcolor{worse}\parbox[c][1.5cm][c]{1.5cm}{\centering \Huge } & \cellcolor{worse}\parbox[c][1.5cm][c]{1.5cm}{\centering \Huge } & \cellcolor{worse}\parbox[c][1.5cm][c]{1.5cm}{\centering \Huge } \\
\hline
\parbox[c][1.5cm][c]{1.5cm}{\centering \Huge \textbf{x5}}  & \cellcolor{better}\parbox[c][1.5cm][c]{1.5cm}{\centering \Huge } & \cellcolor{diaggray}\parbox[c][1.5cm][c]{1.5cm}{\centering \Huge } & \cellcolor{worse}\parbox[c][1.5cm][c]{1.5cm}{\centering \Huge } & \cellcolor{worse}\parbox[c][1.5cm][c]{1.5cm}{\centering \Huge } & \cellcolor{worse}\parbox[c][1.5cm][c]{1.5cm}{\centering \Huge }\\
\hline
\parbox[c][1.5cm][c]{1.5cm}{\centering \Huge \textbf{x10}} & \cellcolor{better}\parbox[c][1.5cm][c]{1.5cm}{\centering \Huge } & \cellcolor{better}\parbox[c][1.5cm][c]{1.5cm}{\centering \Huge } & \cellcolor{diaggray}\parbox[c][1.5cm][c]{1.5cm}{\centering \Huge } & \cellcolor{worse}\parbox[c][1.5cm][c]{1.5cm}{\centering \Huge } & \cellcolor{neutral}\parbox[c][1.5cm][c]{1.5cm}{\centering \Huge } \\
\hline
\parbox[c][1.5cm][c]{1.5cm}{\centering \Huge \textbf{x20}} & \cellcolor{better}\parbox[c][1.5cm][c]{1.5cm}{\centering \Huge } & \cellcolor{better}\parbox[c][1.5cm][c]{1.5cm}{\centering \Huge } & \cellcolor{better}\parbox[c][1.5cm][c]{1.5cm}{\centering \Huge ~} & \cellcolor{diaggray}\parbox[c][1.5cm][c]{1.5cm}{\centering \Huge } & \cellcolor{neutral}\parbox[c][1.5cm][c]{1.5cm}{\centering \Huge } \\
\hline
\parbox[c][1.5cm][c]{1.5cm}{\centering \Huge \textbf{x30}} & \cellcolor{better}\parbox[c][1.5cm][c]{1.5cm}{\centering \Huge } & \cellcolor{better}\parbox[c][1.5cm][c]{1.5cm}{\centering \Huge } & \cellcolor{neutral}\parbox[c][1.5cm][c]{1.5cm}{\centering \Huge } & \cellcolor{neutral}\parbox[c][1.5cm][c]{1.5cm}{\centering \Huge } & \cellcolor{diaggray}\parbox[c][1.5cm][c]{1.5cm}{\centering \Huge }\\
\hline
\end{tabular}
&
\begin{tabular}{|p{1.5cm}|p{1.5cm}|p{1.5cm}|p{1.5cm}|p{1.5cm}|p{1.5cm}|p{1.5cm}|}
\hline
\diagbox[width=1.5cm, height=1.5cm]{\Huge $i$}{\Huge $j$} 
& \parbox[c][1.5cm][c]{1.5cm}{\centering \Huge \textbf{x1}} 
& \parbox[c][1.5cm][c]{1.5cm}{\centering \Huge \textbf{x5}} 
& \parbox[c][1.5cm][c]{1.5cm}{\centering \Huge \textbf{x10}} 
& \parbox[c][1.5cm][c]{1.5cm}{\centering \Huge \textbf{x20}} 
& \parbox[c][1.5cm][c]{1.5cm}{\centering \Huge \textbf{x30}}  \\
\hline
\parbox[c][1.5cm][c]{1.5cm}{\centering \Huge \textbf{x1}}  & \cellcolor{diaggray}\parbox[c][1.5cm][c]{1.5cm}{\centering \Huge } & \cellcolor{worse}\parbox[c][1.5cm][c]{1.5cm}{\centering \Huge } & \cellcolor{worse}\parbox[c][1.5cm][c]{1.5cm}{\centering \Huge } & \cellcolor{worse}\parbox[c][1.5cm][c]{1.5cm}{\centering \Huge } & \cellcolor{worse}\parbox[c][1.5cm][c]{1.5cm}{\centering \Huge } \\
\hline
\parbox[c][1.5cm][c]{1.5cm}{\centering \Huge \textbf{x5}}  & \cellcolor{better}\parbox[c][1.5cm][c]{1.5cm}{\centering \Huge } & \cellcolor{diaggray}\parbox[c][1.5cm][c]{1.5cm}{\centering \Huge } & \cellcolor{worse}\parbox[c][1.5cm][c]{1.5cm}{\centering \Huge } & \cellcolor{worse}\parbox[c][1.5cm][c]{1.5cm}{\centering \Huge } & \cellcolor{worse}\parbox[c][1.5cm][c]{1.5cm}{\centering \Huge }\\
\hline
\parbox[c][1.5cm][c]{1.5cm}{\centering \Huge \textbf{x10}} & \cellcolor{better}\parbox[c][1.5cm][c]{1.5cm}{\centering \Huge } & \cellcolor{better}\parbox[c][1.5cm][c]{1.5cm}{\centering \Huge } & \cellcolor{diaggray}\parbox[c][1.5cm][c]{1.5cm}{\centering \Huge } & \cellcolor{neutral}\parbox[c][1.5cm][c]{1.5cm}{\centering \Huge } & \cellcolor{worse}\parbox[c][1.5cm][c]{1.5cm}{\centering \Huge } \\
\hline
\parbox[c][1.5cm][c]{1.5cm}{\centering \Huge \textbf{x20}} & \cellcolor{better}\parbox[c][1.5cm][c]{1.5cm}{\centering \Huge } & \cellcolor{better}\parbox[c][1.5cm][c]{1.5cm}{\centering \Huge } & \cellcolor{neutral}\parbox[c][1.5cm][c]{1.5cm}{\centering \Huge ~} & \cellcolor{diaggray}\parbox[c][1.5cm][c]{1.5cm}{\centering \Huge } & \cellcolor{worse}\parbox[c][1.5cm][c]{1.5cm}{\centering \Huge } \\
\hline
\parbox[c][1.5cm][c]{1.5cm}{\centering \Huge \textbf{x30}} & \cellcolor{better}\parbox[c][1.5cm][c]{1.5cm}{\centering \Huge } & \cellcolor{better}\parbox[c][1.5cm][c]{1.5cm}{\centering \Huge } & \cellcolor{better}\parbox[c][1.5cm][c]{1.5cm}{\centering \Huge } & \cellcolor{better}\parbox[c][1.5cm][c]{1.5cm}{\centering \Huge } & \cellcolor{diaggray}\parbox[c][1.5cm][c]{1.5cm}{\centering \Huge }\\
\hline
\end{tabular}

\end{tabular}
}
\begin{center}
\renewcommand{\arraystretch}{1.3}
\setlength{\fboxsep}{0pt} 
\setlength{\fboxrule}{0.5pt} 

\begin{tabularx}{\textwidth}{@{}>{\centering\arraybackslash}m{0.06\textwidth}
                            >{\raggedright\arraybackslash}X
                            >{\centering\arraybackslash}m{0.06\textwidth}
                            >{\raggedright\arraybackslash}X
                            >{\centering\arraybackslash}m{0.06\textwidth}
                            >{\raggedright\arraybackslash}X@{}}
\fcolorbox{black}{white}{\makebox[10pt][c]{\phantom{X}}} & Model $i$ and Model $j$ are not statistically different &
\fcolorbox{black}{worse}{\makebox[10pt][c]{\phantom{X}}} & Model $i$ is statistically worse than Model $j$ &
\fcolorbox{black}{better}{\makebox[10pt][c]{\phantom{X}}} & Model $i$ is statistically better than Model $j$ \\
\end{tabularx}
\end{center}

\begin{center}
\small
\hspace*{0pt}(a) City center \hspace{3.2cm} 
(b) Stadium \hspace{3.2cm} 
(c) Airport\hspace{3.2cm} 
(d) Highway
\end{center}


\caption{%
Statistical comparison of KDE-\blueR{1NN} model variants across four deployment scenarios.
Each subfigure corresponds to a cross-model comparison at increasing augmentation rates.
}
\label{fig:wknn_comparison}
\end{figure*}
The goal of this subsection is to quantify how positioning performance of the proposed KDE-\blueR{1NN} architecture scales with the number of synthetic fingerprints and to test whether performance gains saturate as the augmentation factor increases. 
In order to statistically compare model performance metrics in a pairwise manner across augmentation rates, we evaluate whether the difference $d = e_A - e_B$ between
two model positioning errors $e_A$ and $e_B$ for models $A$ and $B$ is statistically significant \cite{statComp, statComp2}.
We consider two models that are evaluated $N_e$ times, yielding individual positioning errors
$e_{A,i}$ and $e_{B,i}$ for \(i=1,\dots,N_e\).  
We model the sample mean for the two compared models as
\begin{equation}
   \overline{e}_A
= \frac{1}{N_e}\sum_{i=1}^{N_e} e_{A,i}~, \quad\quad\overline{e}_B
= \frac{1}{N_e}\sum_{i=1}^{N_e} e_{B,i}~, 
\end{equation}
and their sample variance as
\begin{equation}
    \hat{\sigma}_A^2
= \frac{\sum_{i=1}^{N_e}\bigl(e_{A,i} - \overline{e}_A\bigr)^{2}}{N_e-1}~,
\quad
\hat{\sigma}_B^2
= \frac{\sum_{i=1}^{N_e}\bigl(e_{B,i} - \overline{e}_B\bigr)^{2}}{N_e-1}~.
\end{equation}
For $N_e \geq 30$, by the Central Limit Theorem \cite{kwak2017central}, $e_A$ and $e_B$ converge in distribution to $\mathcal{N}\!\bigl(\overline{e}_A,\;\hat{\sigma}_A^2/N_e\bigr)$ and $\mathcal{N}\!\bigl(\overline{e}_B,\;\hat{\sigma}_B^2/N_e\bigr)$, respectively. Consequently, $d$ converges in distribution to $\mathcal{N}\!\bigl(\bar{d},\;\hat\sigma_d^2\bigr)$, where $\bar{d} = \overline{e}_A - \overline{e}_B$ and $\hat\sigma_d^2 = (\hat{\sigma}_A^2+\hat{\sigma}_B^2)/N_e$.
For confidence level \(1-\alpha\), the \ac{CI} for $\mathbb{E}[d]$ is determined by
\begin{equation}
\label{eq:ci}
    \mathbb{E}[d] \in \left[\bar{d} - z_{\alpha/2}\,\frac{\hat\sigma_d^2}{\sqrt{N_e}}\,,\bar{d} + z_{\alpha/2}\,\frac{\hat\sigma_d^2}{\sqrt{N_e}}\right]~,
\end{equation}
where \(z_{\alpha/2}\) is the standard Normal quantile.
If the \ac{CI} in~\eqref{eq:ci} contains zero, we conclude that the difference between the two models is not statistically significant.
Otherwise, we conclude that the difference is statistically significant, i.e., that model $A$ outperforms model $B$ if the entire \ac{CI} lies above zero and, conversely, that $B$ outperforms $A$ if it lies below zero.

In order to perform the pairwise comparisons, we followed the methodology described above by performing $N_e=30$ independent runs for each model given a region and an augmentation rate, considering a confidence level of 95\% (i.e., $\alpha = 0.05$).
Fig.~\ref{fig:wknn_comparison} summarizes these pairwise comparisons for each region: rows and columns index augmentation rates \(A\in\{1,5,10,20,30\}\), and each cell indicates whether the difference in the chosen positioning metric between the row- and column-level is statistically positive, negative, or not significant, highlighting both monotonic trends and saturation points.
\blue{These results reveal a clear structural difference between the considered areas. In the city center scenario, which features the highest UE density, we observe that performance saturates starting at moderate augmentation rates ($A=10$). In denser scenarios, such as stadium and airport, the transition from $A=20$ to $A=30$ yields no significant gain, suggesting that the synthetic data has effectively filled the spatial gaps within the radio map.
In contrast, the highway scenario, which features the lowest UE density, exhibits a sustained improvement, with significant gains persisting up to $A=30$. This behavior is driven by the highly sparse nature of \ac{MDT} data in this area.
Consequently, we conclude that the saturation point for positioning accuracy is inversely related to the initial UE density of the training data: sparse topologies (e.g., highway) exhibit sustained performance gains at higher augmentation rates, whereas dense environments rapidly converge to a saturation point.}

\blueR{
\subsection{Computational Complexity}
\label{sec:complexity}
%
To evaluate the scalability of the proposed mobile data augmentation framework, we analyze its computational time complexity and benchmark its empirical execution time against the considered baselines. Notably, unlike the parameterized and learning-based baselines, the proposed KDE-1NN pipeline is entirely non-parametric and training-free, operating directly on the empirical reference database. To ensure a fair comparison focused strictly on the synthesis overhead, our evaluation considers only the sample generation phase, assuming all baseline models have already been trained offline.
}
\blueR{
\subsubsection{Theoretical Complexity} Let $m$ represent the size of the empirical reference database
$\green{\mathbf{R}}$,
$N_c$ the number of \acp{PCI}, and $A \in \{1, 5, 10, 20, 30\}$ the augmentation rate. The generation phase must synthesize $N = A \cdot m$ synthetic data samples. The spatial features generation via Gaussian KDE requires $\mathcal{O}(A \cdot m)$ operations, since drawing each synthetic sample involves uniformly selecting one of the $m$ reference locations and applying a Gaussian perturbation, taking $\mathcal{O}(1)$ time per sample. For the radio features augmentation, the \blueR{1NN} algorithm leverages a KD-Tree spatial index to perform neighbor searches in $\mathcal{O}(A \cdot m \log m)$ time, assuming average-case logarithmic search complexity with a preconstructed spatial index. The subsequent RSRP assignment and log-normal shadowing perturbation scale linearly with the number of cells, requiring $\mathcal{O}(A \cdot m \cdot N_c)$ operations. Thus, the overall generation complexity of the proposed KDE-\blueR{1NN} framework is bounded by $\mathcal{O}(A \cdot m (\log m + N_c))$ under these assumptions.
}
\blueR{
\subsubsection{Empirical Benchmarks}
To quantify this theoretical bound, we evaluate the elapsed wall-clock time required to generate the synthetic datasets at high augmentation rates. Table \ref{tab:complexity} reports the elapsed time (in seconds) for each augmentation strategy evaluated {on an NVIDIA RTX 4090 GPU}, using the city center scenario. The reported times represent the median of 10 runs, including KDE sampling and radio feature inference, while excluding model loading and file I/O operations. The empirical results confirm that the KDE-\ac{1NN} framework maintains the lowest computational footprint among all evaluated methods, achieving robust dataset generation in well under half a second even at $A=30$. In contrast, {\ac{RF} and \ac{ReFlow} take} substantially longer{,} {and the deep generative baselines (\ac{MLP}, \ac{CGAN}, \ac{NF}}) achieve comparable generation times to the proposed method.} \blueR{\ac{GPR} is by far the slowest method{: this follows from its need to evaluate the full covariance matrix over the reference database at prediction time, a cost that grows rapidly with the size of the database and dominates its runtime.} These findings validate the lightweight, low-complexity, and scalable nature of the proposed architecture.}

\begin{table}[t]
\centering
\blueR{
\caption{\blueR{Elapsed time for synthetic data generation across high augmentation rates, measured on the city center scenario.}}
\label{tab:complexity}
\small
\renewcommand{\arraystretch}{1.3}
\begin{tabular}{lccc}
\toprule
\textbf{Model} & \textbf{$A=10$} & \textbf{$A=20$} & \textbf{$A=30$} \\
\midrule
KDE-\blueR{1NN} (Proposed) & 0.06 s & 0.13 s & 0.19 s \\
KDE-CGAN & 0.04 s & 0.08 s & 0.11 s \\
KDE-NF & 0.07 s & 0.13 s & 0.20 s \\
KDE-MLP & 0.05 s & 0.10 s & 0.15 s \\
KDE-RF & 1.21 s & 1.90 s & 2.61 s \\
KDE-ReFlow & 0.59 s & 1.18 s & 1.78 s \\
KDE-GPR (RQ) & 153.60 s & 264.80 s & 446.85 s \\
\bottomrule
\end{tabular}
}
\end{table}

\section{Conclusion}
\label{sec:conclusion}
This paper has presented a practical, modular framework for mobile data augmentation tailored to real-world MDT traces collected by \acp{MNO}. The proposed two-stage architecture decouples spatial sample generation from radio-feature synthesis through \ac{KDE} and \blueR{\ac{1NN}}, delivering an interpretable and computationally lightweight solution that directly leverages operator-held geo-tagged measurements.
\blue{%
Extensive experimental validation on four large-scale operational datasets demonstrates that the proposed pipeline yields consistent accuracy gains, particularly in scenarios characterized by sparse sampling and high mobility. Specifically, our \ac{KDE}-\blueR{\ac{1NN}} framework reduced the median positioning error by approximately $30\%$ in the sparse \emph{highway} scenario and by $23\%$ in the complex \emph{stadium} environment, significantly outperforming more complex deep learning and generative baselines.
Furthermore, our analysis revealed a critical structural dependency between augmentation efficacy and the native density of the environment. We observed that benefits from augmentation saturate more rapidly in dense urban centers than high-mobility and sparse regions. This underscores a key operational insight: augmentation strategies must be dynamically tuned to local data sparsity, as indiscriminate over-sampling in already dense areas yields diminishing returns and may not introduce useful information.
In sum, the proposed \ac{KDE}-\blueR{\ac{1NN}} augmentation offers a simple, interpretable, and deployment-ready path to more accurate outdoor fingerprinting on real \ac{MDT} data, while providing operator-relevant guidance on when additional synthetic samples effectively improve positioning.}

\begingroup
\small
\bibliographystyle{IEEEtran}
\bibliography{references}
\endgroup

\begin{IEEEbiography}[{\includegraphics[width=1in,height=1.25in,clip,keepaspectratio]{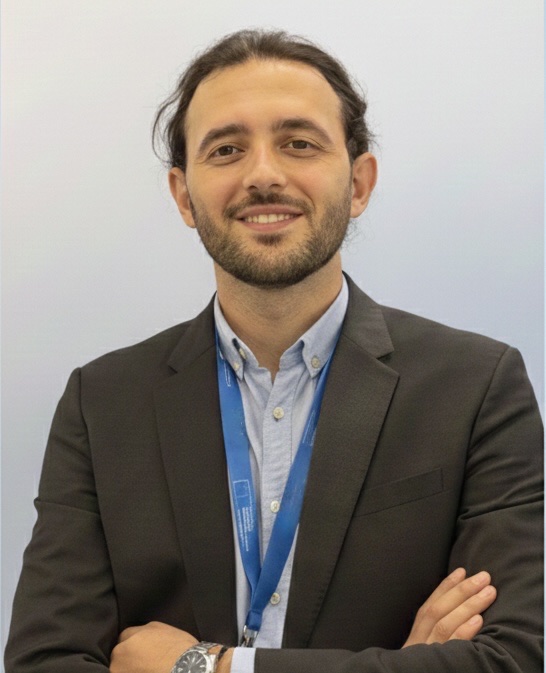}}]{TONY CHAHOUD } (Member, IEEE) is a Researcher at WiLab, the National Laboratory of Wireless Communications of CNIT, Italy. He received the Engineering Degree in Computer and Communications Engineering from the American University of Science and Technology, Beirut, Lebanon, in 2020. In 2024, he received the M.S. degree in Artificial Intelligence from the University of Bologna, Italy. His research interests broadly span machine learning and its applications to wireless communications and Industrial IoT.
\end{IEEEbiography}

\begin{IEEEbiography}[{\includegraphics[width=1in,height=1.25in,clip,keepaspectratio]{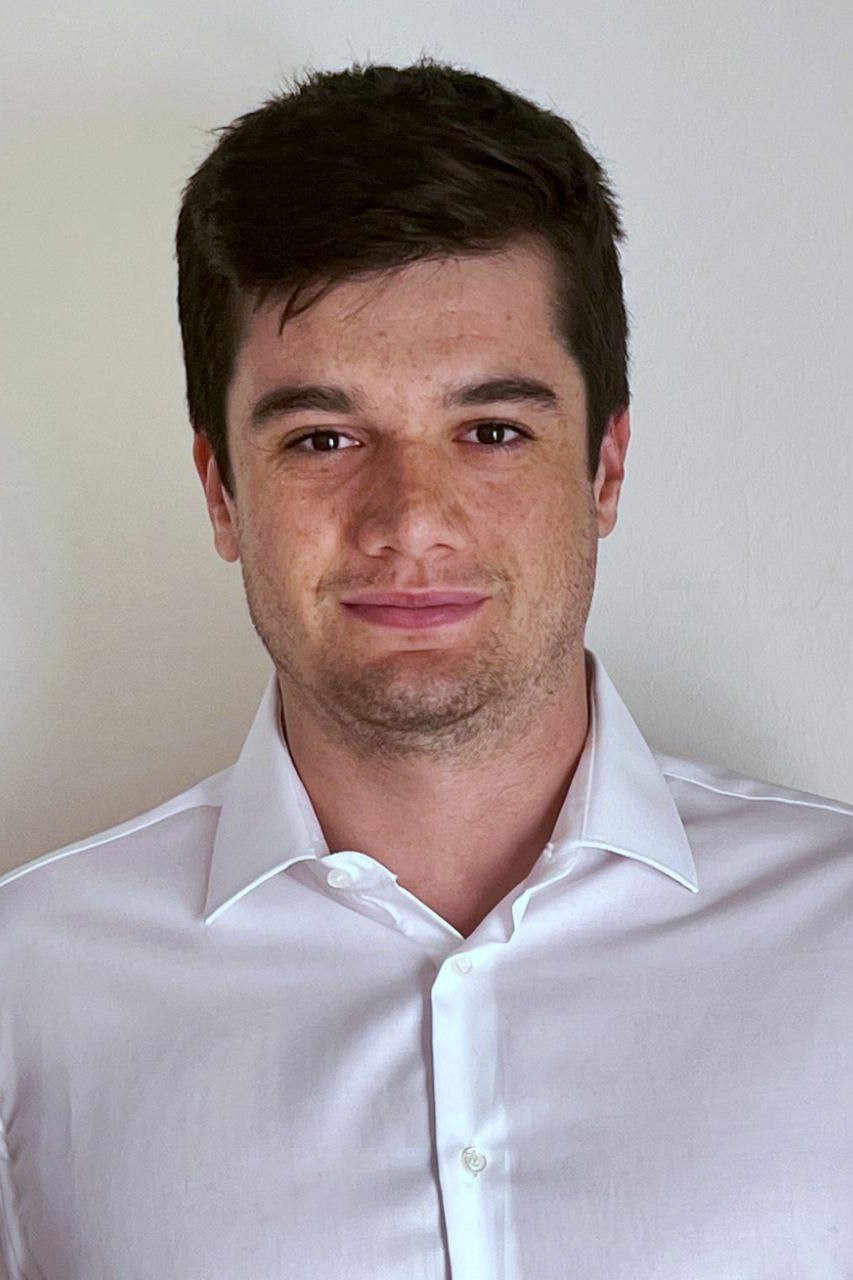}}]{LORENZO MARIO AMOROSA } (Member, IEEE) received the B.S. degree (\textit{summa cum laude}) in Computer Engineering and the M.S. degree (\textit{summa cum laude}) in Artificial Intelligence from the University of Bologna, Italy, in 2019 and 2021, respectively. In 2025, he received the Ph.D. degree in Electronics, Telecommunications and Information Technologies Engineering from the University of Bologna. Currently he is with the Department of Electronic, Information and Electrical Engineering ``Guglielmo Marconi'' as a Postdoctoral Research Fellow at University of Bologna. He is Research Associate at the National Laboratory of Wireless Communications (WiLab) of CNIT (the National, Inter-University Consortium for Telecommunications). His main research interests include decentralized artificial intelligence, cooperative multi-agent systems, and deep learning for wireless communications and IoT.
\end{IEEEbiography}

\begin{IEEEbiography}[{\includegraphics[width=1in,height=1.25in,clip,keepaspectratio]{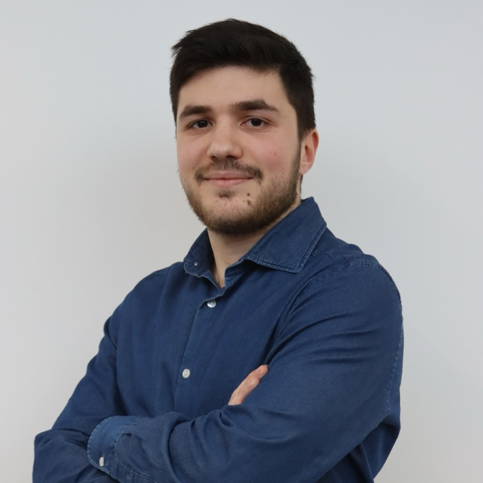}}]{RICCARDO  } is Head of Research at WiLab, the National Laboratory of Wireless Communications of CNIT, where he serves as Sector Leader for Research and Knowledge Transfer. He received his M.Sc. degree in Telecommunications Engineering in 2019 and his Ph.D. in Electronics, Telecommunications and Information Technology in 2023 from the University of Bologna. His research interests include Network Digital Twins, Internet of Things, non-terrestrial networks, and drone communications. He coordinates research activities with industrial partners, focusing on the design and validation of next-generation network architectures, protocols, and proof-of-concepts.
\end{IEEEbiography}

\begin{IEEEbiography}[{\includegraphics[width=1in,height=1.25in,clip,keepaspectratio]{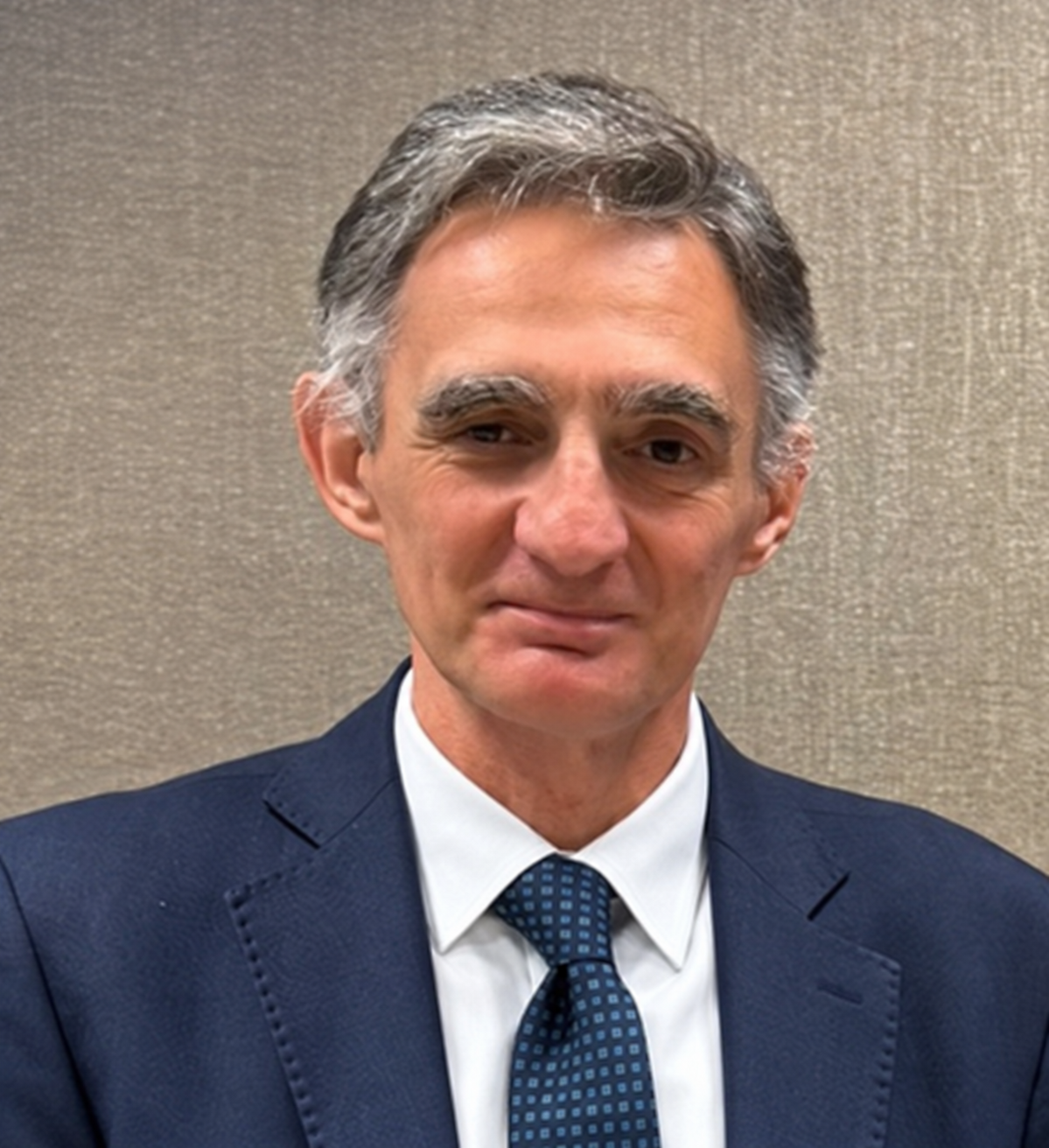}}]{LUCA DE NARDIS } (Member, IEEE) received the Ph.D. degree from the Sapienza University of Rome, Italy, in 2005. He is currently an Associate Professor with the Department of Information Engineering, Electronics and Telecommunications, Sapienza University of Rome. In 2007, he was a Postdoctoral Fellow with the University of California at Berkeley, Berkeley. He authored or co-authored over 150 international peer-reviewed publications. His research interests include wireless positioning systems, cognitive communications and medium access control and routing protocols.
\end{IEEEbiography}

\clearpage

\end{document}